\documentclass[aps,pra,preprint, superscriptaddress,amsmath,amssymb,showpacs]{revtex4-2}
\usepackage[english]{babel}
\usepackage{amssymb,bm}
\usepackage{graphicx}
\usepackage{amsmath}
\usepackage{color}
\usepackage{comment}
\usepackage{multirow}
\usepackage[colorlinks=true,citecolor=blue,urlcolor=blue]{hyperref}
\begin{document}
\begin{titlepage}

\title{Relativistic effects in heavy mesons}
\author{I.V. Obraztsov}\email{ivanqwicliv2@gmail.com}
\author{A.~E.~Bondar}\email{A.E.Bondar@inp.nsk.su}
\author{A.~I.~Milstein}\email{A.I.Milstein@inp.nsk.su}
\affiliation{Budker Institute of Nuclear Physics of SB RAS, 630090 Novosibirsk, Russia}
\affiliation{Novosibirsk State University, 630090 Novosibirsk, Russia}

\date{\today}

\begin{abstract}
We discuss the application of a relativistic potential model to the description of the spectrum and radiative transitions in mesons containing at least one heavy quark ($b$ or $c$). Although the model has a  small number of parameters, it is possible to achieve qualitative agreement with all available experimental data, including those that could not be explained by all previous methods. This demonstrates the importance of taking relativistic effects into account. A remarkable property of the relativistic potential model is that the predictions for meson masses and partial widths of radiative transitions remain finite in the limit of zero light quark mass.	
\end{abstract}

\maketitle
\end{titlepage}

\section{Introduction}
One of the most popular methods for the qualitative study of the spectra of mesons consisting of a heavy quark and an antiquark, and the partial widths of radiative transitions in these systems is the use of nonrelativistic potential models \cite{Bykov:1984}. However, the use of these models to describe systems consisting of one heavy quark ($b$ and $c$) and a light antiquark ($\bar s$, $\bar u$, and $\bar d$) is impossible due to the large relativistic effects related to the motion of the light antiquark. Significant progress in the study of such systems is made in Ref.~\cite{GI1985}, in which all relativistic effects in the spin-independent part of the Hamiltonian is accumulated in the dispersion relation, $m+\bm p^2/2m\to \sqrt{\bm p^2+m^2}$, where $m$ is the quark mass and $\bm p$ is its momentum, $\hbar=c=1$. The spin-dependent part of the Hamiltonian in Ref.~\cite{GI1985} coincides  with the first term of the expansion in $\bm p^2/m^2$ of the corresponding Hamiltonian (the spin part of the Breit Hamiltonian \cite{Breit:1929zz}). However, relativistic corrections to the spin-independent part of the Hamiltonian, which are not related to the dispersion relation, are absent in Ref.~\cite{GI1985}. Although many predictions of the  model \cite{GI1985} and of its various modifications agree well with experiment, some predictions concerning the hierarchy of partial widths of $E1$ and $M1$ radiative transitions, as well as predictions for the values of some hyperfine splittings, clearly contradict experiment. This contradiction was especially pronounced in the $c\bar s$ system with orbital angular momentum $L=0$ and $L=1$. It is due to the presence of strong compensation in the nonrelativistic electric dipole and magnetic moments in such a system \cite{Bondar:2025gsh}.

It is shown in Ref.~\cite{BM2025a} that the use of nonrelativistic electric and magnetic dipole moments in calculating the partial widths of radiative transitions in Ref.~\cite{GI1985} results in a contradiction between the predictions and the experimental data. In Ref.~\cite{BM2025a}, the first relativistic corrections to the Hamiltonian, which contribute to the radiative transition amplitudes, are taken into account, as well as all first-order relativistic corrections in $\bm p^2/m^2$ to the spin-independent part of the Hamiltonian, which determine the meson spectrum. As a result, good agreement is achieved between the predictions and the available experimental data.

Unfortunately, it is impossible to make progress in describing systems consisting of a heavy quark and  $u$ or $d$ antiquark using the Hamiltonian in Ref.~\cite{BM2025a}, since the expansion in $\bm p^2/m^2$ does not work. This difficulty is overcome in Ref.~\cite{BM2025b}, where a relativistic potential model is proposed that is applicable to any masses of light quarks. Using this model, in Ref.~\cite{BM2025b} predictions are obtained  for the partial widths of $M1$ radiative transitions and hyperfine splittings that are in good agreement with experiment.

In this paper, we follow the approach of Ref.~\cite{BM2025b} to analyze the systems $c\bar c$, $b\bar b$, $c\bar s$, $c\bar u$, $c\bar d$, $b\bar{c}$, $b\bar s$, $b\bar u$, and $b\bar d$ with $L=0$ and $L=1$, which are in the lowest-energy states (without radial excitations) for the total spin $S_{tot}=0,\,1$ and the total  angular momentum $\bm J=\bm L+\bm S_{tot}$. Although the number of parameters in the model is small compared to the amount of experimental data, we have managed to achieve qualitative agreement between theory and experiment even in cases where the experimental data are considered counter-intuitive and inexplicable.

Our work is organized as follows. First, we discuss the Hamiltonian of the model, which describes the spectrum of heavy mesons. By heavy mesons we mean mesons containing either a heavy quark and an antiquark, or a heavy quark and a light antiquark. Then we discuss the Hamiltonian responsible for radiative transitions. Next, we find the corresponding wave functions of the mesons and obtain predictions for their mass spectra. Comparison with experimental data  allows us to fix the model parameters. Using the found parameter values, we predict the partial widths of the $E1$ and $M1$ transitions. The conclusion formulates the main results of our work. Some technical details of the calculations are given in the Appendices.

\section{Hamiltonian describing the spectrum of mesons}
As shown in Ref.~\cite{BM2025b}, the Hamiltonian describing the spectrum of heavy mesons in the relativistic potential model is the sum of three terms,
\begin{align}\label{Htot}
	&H=H^{(0)}+\Delta H^{(0)}+\Delta H^{(S)}\,,
\end{align} 
where the leading contribution $H^{(0)}$ reads
\begin{align}\label{H0}
	&H^{(0)}=h_1+h_2+U_{g }(r)+U_{conf }(r)\,,\nonumber\\
	&h_1=\sqrt{m_1^2+\bm p^2},\quad h_2=\sqrt{m_2^2+\bm p^2},\nonumber\\
	&U_{g }(r)=-\dfrac{g}{r}\,,\quad U_{conf }(r)=br+{\cal C}\,.
\end{align} 
Here $m_1$ and $m_2$ are the antiquark and quark masses, $\bm r=\bm r_1-\bm r_2$ and $\bm p_1=-\bm p_2=\bm p$, where $\bm p=-i\bm\nabla$ is the momentum operator, the potential $U_{g}(r)$ is the zeroth component of the Lorentz vector, and $U_{conf}(r)$ is a Lorentz scalar. The model parameters $g$, $b$, $m_1$, $m_2$, and $\cal C$ are found by comparing the predictions for the meson spectra with experimental data. The Hamiltonian $H^{(0)}$ coincides with the main contribution to the Hamiltonian in the Godfrey–Izgur model \cite{GI1985} and similar later models. The constant ${\cal C}$ in the potential $U_{conf }(r)$ affects the overall shift of energy levels and does not affect their difference.

The Hamiltonian $\Delta H^{(0)}$ is a relativistic correction that is independent of the spins $\bm S_1$ and $\bm S_2$ of the antiquark and quark, respectively, see Ref.~\cite{BM2025b}. It contains three contributions,
\begin{align}\label{DH0}
	&\Delta H^{(0)}=\Delta H^{(0)}_{1}+\Delta H^{(0)}_{2}+\Delta H^{(0)}_{3}\,.
\end{align} 	
The terms $\Delta H^{(0)}_{1}$  and $	\Delta H^{(0)}_{2}$ are related to the potential $U_g(r)$,
\begin{align}\label{DH012}
	&\Delta H^{(0)}_{1}=-\dfrac{g}{4}\Bigg[\dfrac{p^i}{h_2}\left(\dfrac{\delta^{ij}}{r}+\dfrac{r^ir^j}{r^3}\right)\dfrac{p^{j}}{h_1}+\dfrac{p^i}{h_1}\left(\dfrac{\delta^{ij}}{r}+\dfrac{r^ir^j}{r^3}\right)\dfrac{p^{j}}{h_2}\Bigg]\,,\nonumber\\
	&\Delta H^{(0)}_{2}=\dfrac{\pi g}{2}\,\left[ \dfrac{1}{h_1}\,\delta(\bm r)\dfrac{1}{h_1}+\dfrac{1}{h_2}\,\delta(\bm r)\dfrac{1}{h_2}\right]\,,
	\end{align} 
where $\delta(\bm r)$ is the Dirac delta function. The contribution $\Delta H^{(0)}_{2}$ (the so-called Darwin term), although independent of spin. Note that it exists only for particles with spin.

Finally, the contribution $\Delta H^{(0)}_{3}$ is related to the potential $U_{conf}(r)$,
 \begin{align}\label{DH03}
	&\Delta H^{(0)}_{3}=\dfrac{b}{2}\,\Bigg[ \dfrac{1}{h_1}\, \left(\dfrac{1}{2r} -p^irp^i\right)\dfrac{1}{h_1}+ \dfrac{1}{h_2}\, \left(\dfrac{1}{2r} -p^irp^i\right) \dfrac{1}{h_2}\Bigg]\,.
\end{align}
It is clear that the potentials $U_{g}(r)$ and $U_{conf}(r)$ contribute differently to the $\Delta H^{(0)}$ correction. Note that the $\Delta H^{(0)}$ correction is absent in the Godfrey-Izgur model \cite{GI1985}. Replacing $h_{1,2}\to m_{1,2}$ in Eq.~\eqref{DH012}, we obtain the result which coincides with the corresponding contribution to the Breit Hamiltonian in quantum electrodynamics (see, e.g., \cite{Pilkuhn:1979ps, Lee:2001wwa}).

The spin-dependent relativistic correction $\Delta H^{(S)}$ consists of two terms,
\begin{align}\label{DHS}
	&\Delta H^{(S)}=\Delta H^{(S)}_{1}+\Delta H^{(S)}_{2}\,,
\end{align}
where $\Delta H^{(S)}_{1}$ depends linearly on the spin operators and $\Delta H^{(S)}_{2}$ depends quadratically. For $\Delta H^{(S)}_{1}$ we have
\begin{align} \label{DHS1}
	&\Delta H^{(S)}_{1}=\dfrac{1}{2h_1}\left(\dfrac{g}{r^3}-\dfrac{b}{r}\right)(\bm L\cdot \bm S_1)\dfrac{1}{h_1}+\dfrac{1}{2h_2}\left(\dfrac{g}{r^3}-\dfrac{b}{r}\right)(\bm L\cdot \bm S_2)\dfrac{1}{h_2}\nonumber\\
	&+\dfrac{g}{2h_1}\dfrac{(\bm L\cdot\bm S_{tot})}{r^3}\dfrac{1}{h_2}+\dfrac{g}{2h_2}\dfrac{(\bm L\cdot\bm S_{tot})}{r^3}\dfrac{1}{h_1}\,,
\end{align} 
where $\bm L=[\bm r\times\bm p]$ is the angular momentum operator and $\bm S_{tot}=\bm S_{1}+\bm S_{2}$.

The spin-quadratic contribution $\Delta H^{(S)}_{2}$ is
\begin{align}\label{DHS2}
	&\Delta H^{(S)}_{2}=\dfrac{g}{2}\Bigg[\dfrac{1}{h_1}{\cal P}\dfrac{1}{h_2}+\dfrac{1}{h_2}{\cal P}\dfrac{1}{h_1}\Bigg]\,,\nonumber\\
	&{\cal P}=\dfrac{8\pi }{3}\delta(\bm r)(\bm S_1\cdot\bm S_2)+\dfrac{1}{r^3}\Big[3(\bm S_1\cdot\bm n)(\bm S_2\cdot \bm n)-(\bm S_1\cdot\bm S_2)\Big]\,,
\end{align}
where $\bm n=\bm r/r$. Below, eigenvalues of the Hamiltonian $H$ in Eq.~\eqref{Htot}, corresponding to the certain angular momentum $L$ and the total angular momentum $J$,
are denoted by ${\cal E}^J_L$.

\section{Wave functions and spectrum of mesons}
For a certain angular momentum $L$, total spin $S_{tot}$, total angular momentum $J$, and total projection $J_z=\mu$, the eigenfunctions of the Hamiltonian $H^{(0)}$ in Eq.~\eqref{H0} have the form $\Phi^{J,\mu}_{L,S_{tot}}(\bm p)$ in the momentum representation and $\Psi^{J,\mu}_{L,S_{tot}}(\bm r)$ in the coordinate representation,
\begin{align}\label{WFp}
	&\Phi^{J,\mu}_{L,S_{tot}}(\bm p)=i^{-L}\phi_L(p)\, \chi^{J,\mu}_{L,S_{tot}}(\bm n_p)\,,\nonumber \\
	&\Psi^{J,\mu}_{L,S_{tot}}(\bm r)=\psi_L(r)\, \chi^{J,\mu}_{L,S_{tot}}(\bm n)\,,\nonumber \\
	&\chi^{J,\mu}_{L,S_{tot}}(\bm n_p)=\sum_{\mu'}C^{J\mu}_{L\mu',S_{tot} (\mu-\mu')}\,Y_{L,\mu'}(\bm n_p)\,|S_{tot},\mu-\mu'\rangle\,.
\end{align} 
Here $Y_{L,\mu}(\bm n)$ are spherical functions, $C^{J\mu}_{L\mu',S_{tot} (\mu-\mu')}$ are the Clebsch-Gordan coefficients, $\bm n_p=\bm p/p$, $\phi_L(p)$ and $\psi_L(r)$ are the radial parts of the corresponding wave functions. The eigenvalues $E_L^{(0)}$ of the Hamiltonian $H^{(0)}$ are independent of $J$ and $S_{tot}$. To calculate the matrix elements we also need the functions
 \begin{align}\label{Xi}
	&\Xi^{J,\mu}_{L,S_{tot}}(\bm r,m)=\int\dfrac{d^3p}{(2\pi)^3}\,\dfrac{\Phi^{J,\mu}_{L,S_{tot}}(\bm p)}{\sqrt{m^2+\bm p^2}}\,e^{i\bm p\cdot\bm r}=\xi_L(r,m)\, \chi^{J,\mu}_{L,S_{tot}}(\bm n)\,.
\end{align}

It is important that the operator $H^{(0)}$ is integral in both the coordinate and momentum representations, and the exact determination of the energies  and wave functions for this Hamiltonian is a  hard problem. The use of the variational method significantly simplifies the calculations \cite{BM2025a, BM2025b}. In this paper, we are interested in the ground radial states of mesons with $L=0$, $L=1$ and different $J$. Small admixtures of states with $L=2$ and $L=3$, respectively, to states with $L=0$ and $L=1$ due to the Hamiltonian $\Delta H^{(0)}_1$ in Eq.~\eqref{DH012} are not considered in our work.

\subsection{Mesons with $L=0$}
For $L=0$ we represent the variational wave function in the form \cite{BM2025a,BM2025b}
\begin{align}\label{WF0}
	&\phi_0( p)=\sqrt{N_0}\,\exp\left[-\dfrac{p^2}{2\omega_0^2}\right]\,,\quad N_0=\dfrac{32\pi^{5/2}}{\omega_0^3}\,,\nonumber\\
	&\psi_0(r)=\sqrt{{\cal N}_0}\,\exp\left[-\dfrac{\omega_0^2 r^2}{2}\right]\,,\quad {\cal N}_0=\dfrac{4\omega_0^3}{\sqrt{\pi}}\,,\nonumber\\
	&\chi^{S_{tot},\mu}_{0,S_{tot}}(\bm n)=\dfrac{1}{\sqrt{4\pi}}\,|S_{tot},\mu\rangle\,,
	\end{align}
where $\omega_0$ is the variational parameter. The average value $E^{(0)}_0$ of the Hamiltonian $H^{(0)}$ over this wave function is
\begin{align}\label{E0}
	&E^{(0)}_0=\omega_0[\gamma_{10}\,{\cal K}_1(\gamma_{10}/2)+\gamma_{20}{\cal K}_1(\gamma_{20}/2)]+
	\dfrac{2}{\sqrt{\pi}}\left[\dfrac{b}{\omega_0}-g\omega_0\right]\,,\nonumber\\
	&\gamma_{10}=\dfrac{m_1^2}{\omega_0^2}\,,\quad \gamma_{20}=\dfrac{m_2^2}{\omega_0^2}\,.
\end{align}
Here and below we use the notation
\begin{align}\label{kalK}
	&{\cal K}_n(x)=\dfrac{1}{\sqrt{\pi}}\,e^xK_n(x)\,,
\end{align}
where $K_n(x)$ is the  modified Bessel function of the third kind. The value of the variational parameter $\omega_0$ is determined by minimizing the value $E^{(0)}_0$. Using variational functions, we obtain for the function $\xi_L(r,m)$ in Eq.~\eqref{Xi} at $L=0$ 
\begin{align}\label{Xi0}
	&\xi_0(r,m)=\dfrac{\sqrt{2{\cal N}_0}}{\omega_0\sqrt{\pi}}\int_0^\infty \dfrac{ds}{(1+s^2)^{3/2}}\,
	\exp\left[-\dfrac{m^2s^2}{2\omega_0^2}-\dfrac{\omega_0^2r^2}{2(1+s^2)}\right]\,.
\end{align} 
Then we find the spin-independent correction $\Delta E^{(0)}_0$ to the energy level,
\begin{align}\label{DE00}
	&\Delta E^{(0)}_0=\Delta E^{(0)}_{01}+\Delta E^{(0)}_{02}+\Delta E^{(0)}_{03}\,,\nonumber\\
	&\Delta E^{(0)}_{01}=	-\dfrac{g\omega_0}{\sqrt{\pi}}\int_0^\infty dy\,y e^{-y}\,{\cal K}_0(z_{10}){\cal K}_0(z_{20})\,,\nonumber\\ 
	&\Delta E^{(0)}_{02}=\dfrac{g\,\omega_0}{16\sqrt{\pi}}\left\{\gamma_{10}^2\left[{\cal K}_1\left(\gamma_{10}/4\right)-{\cal K}_0\left(\gamma_{10}/4\right)\right]^2   +(\gamma_{10}\rightarrow \gamma_{20}) \right\}\,,\nonumber\\
	&\Delta E^{(0)}_{03}= \dfrac{b}{8\omega_0\sqrt{\pi}}\int_0^\infty dy\,e^{-y}\,\Bigg\{2{\cal K}_0^2(z_{10})-y^2[{\cal K}_0(z_{10})+{\cal K}_1(z_{10})]^2 + ( z_{10}\rightarrow z_{20})\Bigg\}\,,\nonumber\\
	&z_{10}=\dfrac{1}{4}\left(y+\gamma_{10}\right)\,,\quad	z_{20}=\dfrac{1}{4}\left(y+\gamma_{20}\right)\,.
\end{align}
For $L=0$, the average value $\Delta E^{(S)}_{01}$ of the term $\Delta H^{(S)}_{1}$ vanishes, and the term $\Delta H^{(S)}_{2}$ yields an energy shift
\begin{align} \label{DE0S}
	&\Delta E^{(S)}_{0}=-\dfrac{3}{4}\Omega\,\delta_{S_{tot},\,0}+\dfrac{1}{4}\Omega\,\delta_{S_{tot},\,1}\,.
\end{align}
Here $\delta_{n_1,n_2}$ is the Kronecker symbol and $\Omega$ is the value of hyperfine splitting (the energy difference between states with $S_{tot}=1$ and $S_{tot}=0$),
\begin{align}\label{Omfvar}
	&\Omega=\dfrac{g\,\omega_0}{3\sqrt{\pi}}\,\gamma_{10}\gamma_{20}\,\left[{\cal K}_1\left(\gamma_{10}/4\right)-{\cal K}_0\left(\gamma_{10}/4\right)\right]\left[{\cal K}_1\left(\gamma_{20}/4\right)-{\cal K}_0\left(\gamma_{20}/4\right)\right]\,.
\end{align} 
Finally, for $J=1$ and $J=0$ at $L=0$, we find the energies with account for the corrections, 
\begin{align}\label{E0Jtot}
	&{\cal E}_0^1=E^{(0)}_0+\Delta E^{(0)}_0+\dfrac{1}{4}\Omega\,,\quad {\cal E}_0^0=E^{(0)}_0+\Delta E^{(0)}_0-\dfrac{3}{4}\Omega\,.
\end{align}
The quantities  in these formulas are defined in Eqs.~\eqref{E0}, \eqref{DE00}, and \eqref{Omfvar}.

\subsection{Mesons with $L=1$}
If $L=1$ and $S_{tot}=1$, then the total momentum is $J=0,\,1,\,2$, and if $L=1$ and $S_{tot}=0$, then $J=1$. The variational wave functions for $L=1$ and certain $S_{tot}$, $J$ and $J_z=\mu$ can be written as
\begin{align} \label{WF1}
	&\phi_{1}(p)=-i\sqrt{N_1}\,p\,\exp\left[-\dfrac{p^2}{2\omega_1^2}\right]\,,\quad N_1=\dfrac{8\,(2\pi)^3}{3\sqrt{\pi}\omega_1^5}\,,\nonumber\\
	&\psi_1(r)=\sqrt{{\cal N}_1}\,r\,\exp\left[-\dfrac{\omega_1^2r^2}{2}\right]\,,\quad {\cal N}_1=\dfrac{8}{3\sqrt{\pi}}\,\omega_1^5\,,\nonumber\\
	&\chi^{J,\mu}_{1,S_{tot}}(\bm n)=i\sqrt{\dfrac{3}{4\pi}}\sum_{\mu'}C^{J\mu}_{1\mu',S_{tot} (\mu-\mu')}\,(\bm e_{\mu'}\cdot\bm n_p)\,|S_{tot},\mu-\mu'\rangle\,,\nonumber\\
	&\bm e_0=\bm e_z\,,\quad \bm e_1=-\dfrac{(\bm e_x+i\bm e_y)}{\sqrt{2}}\,,\quad \bm e_{-1}=\dfrac{(\bm e_x-i\bm e_y)}{\sqrt{2}}\,.
\end{align} 
Here we have used the relation
$$Y_{1,\mu}(\bm n)=i\sqrt{\dfrac{3}{4\pi}}(\bm e_{\mu}\cdot\bm n)\,.$$
For the function $\xi_{1}(r,m)$ in Eq.~\eqref{Xi}, we find
\begin{align}\label{Xi1}
	&\xi_{1}(r,m)=\dfrac{\sqrt{{2\cal N}_1}}{\omega_1\sqrt{\pi}}\int_0^\infty \dfrac{r\,ds}{(1+s^2)^{5/2}}\,\exp\left[-\dfrac{m^2s^2}{2\omega_1^2}-\dfrac{\omega_1^2r^2}{2(1+s^2)}\right]\,.
\end{align} 
The average value $E^{(0)}_1$ of the Hamiltonian $H^{(0)}$ over the variational function in Eq.~\eqref{WF1} is 
\begin{align}\label{E1}
	&E^{(0)}_1=\omega_1\,[{\cal F}(\gamma_{11}/2)  +{\cal F}(\gamma_{21}/2)] +\dfrac{4}{3\sqrt{\pi}}\left(\dfrac{2b}{\omega_1}-g\omega_1\right)\,,\nonumber\\
	&{\cal F}(x)=\dfrac{4}{3}x\,[x {\cal K}_0(x)+(2-x){\cal K}_1(x)]\,,\nonumber\\
	&\gamma_{11}=\dfrac{m_1^2}{\omega_1^2}\,,\quad \gamma_{21}=\dfrac{m_2^2}{\omega_1^2}\,.
\end{align}
The value of the variational parameter $\omega_1$ is determined by minimizing the value $E^{(0)}_1$.
The correction $\Delta E^{(0)}_1=\Delta E^{(0)}_{11}+\Delta E^{(0)}_{12}+\Delta E^{(0)}_{13}$, which corresponds to the term $\Delta H^{(0)}$, is independent of $J$. We have
\begin{align}\label{DE10}
	&\Delta E^{(0)}_{11}=-\dfrac{2\omega_1g}{3\sqrt{\pi}}\int_0^\infty dy\,y^2 e^{-y}{\cal K}_0(z_{11})K_0(z_{21})\,,\nonumber\\
	&z_{11}=\dfrac{1}{4}\left(y+\gamma_{11}\right)\,,\quad z_{21}=\dfrac{1}{4}\left(y+\gamma_{21}\right)\,,\nonumber\\
	&\Delta E^{(0)}_{12}=0\,,\nonumber\\
	&\Delta E^{(0)}_{13}=\dfrac{b}{2\omega_1\sqrt{\pi}}\int_0^\infty dy\,ye^{-y}\left[{\cal K}^2_0(z_{11})-\dfrac{y^2}{6}[{\cal K}_0(z_{11})+{\cal K}_1(z_{11})]^2+(z_{11}\rightarrow z_{21})\right]\,.
\end{align}

Let us now proceed to the calculation of the corrections $\Delta E^{(S)}_{1J}$ and $\Delta E^{(S)}_{2J}$, which depend on $J$ and correspond to the terms $\Delta H^{(S)}_{1}$ and $\Delta H^{(S)}_{2}$.
\subsubsection{The cases $J=0$ and $J=2$}
The correction $\Delta E^{(S)}_{12}$ has the form  
\begin{align}\label{DE12S1}
	&\Delta E^{(S)}_{12}=\dfrac{2\omega_1}{3\sqrt{\pi}}\int_0^\infty dy\,e^{-y} \Bigg\{g\Bigg[z_{11}^2[{\cal K}_1(z_{11})-{\cal K}_0(z_{11})]^2+z_{21}^2[{\cal K}_1(z_{21})-{\cal K}_0(z_{21})]^2\nonumber\\
	&+4z_{11} z_{21}[{\cal K}_1(z_{11})-{\cal K}_0(z_{11})] [{\cal K}_1(z_{21})-{\cal K}_0(z_{21})]\Bigg]-\dfrac{by}{4\omega_1^2}\left[{\cal K}_0^2(z_{11})+{\cal K}_0^2(z_{21})\right]\Bigg\}\,.
\end{align}
For $L=1$ and $S_{tot}=1$ the relation holds
$$\langle J, L,S_{tot}|\bm L\cdot\bm S_{tot}|J,L,S_{tot}\rangle =\delta_{J,2}-2\delta_{J,0}-\delta_{J,1}\,.$$
Therefore, the correction $\Delta E^{(S)}_{10}$ reads
\begin{align}\label{DE1S10tot}
	&\Delta E^{(S)}_{10}=-2\Delta E^{(S)}_{12}\,. 
\end{align}
The  correction $\Delta E^{(S)}_{22}$ is
\begin{align}\label{DE12S2}
	&\Delta E^{(S)}_{22}=-\dfrac{4\omega_1 g}{15\sqrt{\pi}}\int_0^\infty dy\,e^{-y} z_{11} z_{21}[{\cal K}_1(z_{11})-{\cal K}_0(z_{11})] [{\cal K}_1(z_{21})-{\cal K}_0(z_{21})]\,.
\end{align}
We have used the relation valid for $L=1$ and $S_{tot}=1$,
$$\langle J, L,S_{tot}| [3(\bm S_{tot}\cdot\bm n)^2-\bm S_{tot}^2] |J,L,S_{tot}\rangle=-\dfrac{1}{5}[\delta_{J,2}+10\delta_{J,0}-5\delta_{J,1}]\,.$$
Therefore, we have for the correction $\Delta E^{(S)}_{20}$,
\begin{align}\label{DE1S10}
	&\Delta E^{(S)}_{20}=10\,\Delta E^{(S)}_{22}\,. 
\end{align}
Finally, we have  
\begin{align}\label{E12tot}
	&{\cal E}^2_1=E^{(0)}_1+\Delta E^{(0)}_1+\Delta E^{(S)}_{12}+\Delta E^{(S)}_{22}\,,
\end{align}
for the energy of the state with  $L=1$, $S_{tot}=1$, and $J=2$, and 
\begin{align}\label{E10tot}
	&{\cal E}^0_1=E^{(0)}_1+\Delta E^{(0)}_1-2\Delta E^{(S)}_{12}+10\Delta E^{(S)}_{22}\,,
\end{align}
for the energy of the state with $L=1$, $S_{tot}=1$ and $J=0$. The quantities in these formulas are given in Eqs.~\eqref{E1}, \eqref{DE10}, \eqref{DE12S1}, and \eqref{DE12S2}. 

\subsubsection{The case $J=1$}
The case $J=1$ requires special consideration, since there are two such states, with $S_{tot}=0$ and $S_{tot}=1$. The term $\Delta H^{(S)}_{1}$ in Eq.~\eqref{DHS1} does not commute with the operator $\bm S_{tot}$, so that the states with a certain energy, $\Psi_1(\bm r)$ and $\Psi_1'(\bm r)$, are a superposition of states $\Psi_{1,0}^{1\mu}(\bm r)$ and $\Psi_{1,1}^{1\mu}(\bm r)$,
\begin{align}\label{psi12}
	&\Psi_1(\bm r)=c_1\Psi_{1,0}^{1\mu}(\bm r)+c_2\Psi_{1,1}^{1\mu}(\bm r)\,,\quad \Psi_1'=c_2\Psi_{1,0}^{1\mu}(\bm r)-c_1\Psi_{1,1}^{1\mu}(\bm r)\,.
\end{align} 
The corresponding energies ${\cal E}_1^1$, ${\cal E}_1^{1'}$ and the coefficients $c_{1,2}$ are found by solving the secular equation \cite{LLQM},
\begin{align}\label{seceq}
	&{\cal E}_{1}^1=E^{(0)}_1+\Delta E^{(0)}_1+\Delta{\cal E}_{1}\,,\quad {\cal E}_{1}^{1'}=E^{(0)}_1+\Delta E^{(0)}_1+\Delta{\cal E}_{1}'\,,\nonumber\\
	&\Delta{\cal E}_{1}=\dfrac{(\Delta H^{(S)})_{11}+(\Delta H^{(S)})_{22}}{2}-\gamma\,,\quad \Delta{\cal E}_{1}'=\dfrac{(\Delta H^{(S)})_{11}+(\Delta H^{(S)})_{22}}{2}+\gamma\,,\nonumber\\
	&\gamma=\sqrt{\dfrac{[(\Delta H^{(S)})_{11}-(\Delta H^{(S)})_{22}]^2}{4}+[(\Delta H^{(S)})_{12}]^2}\,,\nonumber\\
	&c_1=-\mbox{sgn}[(\Delta H)_{12}]\,\sqrt{\dfrac{\Delta {\cal E}_1'}{2\gamma}}\,,\quad c_2=\sqrt{\dfrac{|\Delta{\cal E} _1|}{2\gamma}}\,.
\end{align} 
An important characteristic of the states is the mixing angle $\varphi$, determined by the relation $\varphi=\arctan(c_1/c_2)$.

In the limit $m_2\to\infty$, the Hamiltonian $\Delta H^{(S)}$ reads
\begin{align}\label{DHSinf}
	&\Delta H^{(S)}=\dfrac{1}{2h_1}\left(\dfrac{g}{r^3}-\dfrac{b}{r}\right)\dfrac{1}{h_1}(\bm L\cdot \bm S_1)\,.
\end{align}	
	Using this expression we obtain
\begin{align}\label{DHSinf1}
	&c_1=-\sigma \sqrt{\dfrac{3-\sigma}{6}}\,,\quad c_2= \sqrt{\dfrac{3+\sigma}{6}}\,, \nonumber\\
	&\sigma=\mbox{sgn}\left\langle \Psi_{1,0}^{1\mu}\left| \dfrac{1}{2h_1}\left(\dfrac{g}{r^3}-\dfrac{b}{r}\right)\dfrac{1}{h_1}\right|\Psi_{1,0}^{1\mu}\right\rangle=\mbox{sgn}\left\langle \Psi_{1,1}^{1\mu}\left|\dfrac{1}{2h_1}\left(\dfrac{g}{r^3}-\dfrac{b}{r}\right)\dfrac{1}{h_1}\right|\Psi_{1,1}^{1\mu}\right\rangle\,.
\end{align}	
Therefor,
\begin{equation}\label{DHSinf2}
	\begin{cases}
		c_1=-\sqrt{\dfrac{1}{3}}\,,\quad c_2= \sqrt{\dfrac{2}{3}}\,, \quad \varphi=-\arctan(1/\sqrt{2})=-35.3^\circ \quad &\mbox{for}\,\sigma=1 \,,\vspace{2mm}\\
		c_1=\sqrt{\dfrac{2}{3}}\,,\quad c_2= \sqrt{\dfrac{1}{3}}\,, \quad \varphi=\arctan(\sqrt{2})=54.7^\circ \quad &\mbox{for}\,\sigma=-1\,.
	\end{cases}
\end{equation}		
The mixing angle $\varphi$ in the infinitely heavy quark approximation is discussed in Ref.~\cite{Matsuki:2010zy}. In the limit $m_2\to\infty$ it is convenient to introduce the operator $\bm j_1=\bm L+\bm S_1$. Then, the probabilities $W_{1/2}$ and $W_{3/2}$ to find $j_1=1/2$ and $j_1=3/2$ in the system, corresponding to the wave function $\Psi_1'$, are 
\begin{align}\label{DHSinf3}
	&W_{1/2}=\dfrac{1}{3}(\sqrt{2}c_1+c_2)^2=\delta_{\sigma,-1}\,,\quad W_{3/2}=1-W_{1/2}=\delta_{\sigma,1}\,.
\end{align}

For a finite value of a heavy quark mass, in our model we have
\begin{align}\label{DH111222}
	&(\Delta H^{(S)})_{11}=0\,,\quad (\Delta H^{(S)})_{22}=-\Delta E^{(S)}_{12}-5\,\Delta E^{(S)}_{22}\,,\nonumber\\
	& (\Delta H^{(S)})_{12}=\dfrac{2\sqrt{2}\,\omega_1}{3\sqrt{\pi}}\int_0^\infty dy\,e^{-y} \Bigg\{gz_{11}^2[{\cal K}_1(z_{11})-{\cal K}_0(z_{11})]^2\nonumber\\
	&-gz_{21}^2[{\cal K}_1(z_{21})-{\cal K}_0(z_{21})]^2-\dfrac{b y}{4\omega_1^2}\Big[{\cal K}_0^2(z_{11})-{\cal K}_0^2(z_{21})\Big]\Bigg\}\,.
\end{align}
It is evident that ${\cal E}_{1}<{\cal E}_{1}'$.

\section{Hamiltonian describing radiative transitions}
The Hamiltonian $H_{rad}$ describing single-photon transitions in the relativistic potential model is easy to obtain using the results of Refs.~\cite{BM2025a} and \cite{BM2025b}. It has the form
\begin{align}\label{Hrad}
	&H_{rad}=G^{(0)}+\dfrac{1}{2}\bm S_{tot}\cdot\bm G^{(S)}+\dfrac{1}{2}\bm \Sigma\cdot\bm G^{(\Sigma)}\,,
\end{align} 
where $\bm\Sigma=\bm S_1-\bm S_2$. It is convenient to represent the contribution $G^{(0)}$ as a sum $G^{(0)}=G^{(0)}_{1}+G^{(0)}_{2}+G^{(0)}_{3}$, where
\begin{align}\label{G0}
	&G^{(0)}_{1}= -\dfrac{1}{2}\left\{\dfrac{1}{h_1},e_1\bm A_1\cdot\bm p\right\}+\dfrac{1}{2}\left\{\dfrac{1}{h_2},e_2\bm A_2\cdot\bm p\right\}\,,\nonumber\\
	&G^{(0)}_{2}=\dfrac{g}{4}	\Bigg[\dfrac{p^j}{h_2}\left(\dfrac{\delta^{ij}}{r}+\dfrac{r^ir^j}{r^3}\right)e_1A_1^i\dfrac{1}{h_1}+\dfrac{1}{h_1}e_1A_1^i\left(\dfrac{\delta^{ij}}{r}+\dfrac{r^ir^j}{r^3}\right)\dfrac{p^j}{h_2}\nonumber\\
	&-(h_1\leftrightarrow h_2,\, e_1\to e_2,\,A_1\to A_2)\Bigg]\,,\nonumber\\
	&G^{(0)}_{3}=\dfrac{b}{2}\Bigg[\dfrac{1}{h_1}\{ r\,,e_1\bm A_1\cdot\bm p\}\dfrac{1}{h_1}-\dfrac{1}{h_2}\{ r\,,e_2\bm A_2\cdot\bm p\}\dfrac{1}{h_2}\Bigg]\,.
\end{align} 
Here $\{A,B\}=AB+BA$, $e_1$ is the antiquark charge, $e_2$ is the quark charge, and the vector potential $\bm A_i$ of the electromagnetic wave in the final state is 
\begin{align}\label{A}
	&\bm A_i=\bm A_W^* e^{-i\bm k\cdot\bm r_i}\,,\quad \bm r_1=\dfrac{m_2}{M}\bm r\,,\quad \bm r_2=-\dfrac{m_1}{M}\bm r\,,\quad M=m_1+m_2\,,
\end{align}
where $\bm A_W^*$ is the polarization vector perpendicular to the photon momentum $\bm k$.

The operator $\bm G^{(S)}$ in Eq.~\eqref{Hrad} is a sum $\bm G^{(S)}=\bm G^{(S)}_{1}+\bm G^{(S)}_{2}+\bm G^{(S)}_{3}$, where
\begin{align}\label{Gtot}
	&\bm G^{(S)}_{1}= \dfrac{i}{2}\left\{\dfrac{1}{h_1},[\bm k\times e_1\bm A_1]\right\}+\dfrac{i}{2}\left\{\dfrac{1}{h_2},[\bm k\times e_2\bm A_2]\right\}\,,\nonumber\\
	&\bm G^{(S)}_{2}= -\dfrac{1}{2h_1}\left(\dfrac{g}{r^3}-\dfrac{b}{r}\right)[\bm r\times e_1\bm A_1 ]\dfrac{1}{h_1}	+\dfrac{1}{2h_2}\left(\dfrac{g}{r^3}-\dfrac{b}{r}\right)[\bm r\times e_2\bm A_2 ]\dfrac{1}{h_2}\nonumber\\
	&-\dfrac{1}{2h_1}\dfrac{g}{r^3}[\bm r\times( e_1\bm A_1-e_2\bm A_2) ]\dfrac{1}{h_2}-\dfrac{1}{2h_2}\dfrac{g}{r^3}[\bm r\times( e_1\bm A_1-e_2\bm A_2) ]\dfrac{1}{h_1}\,,\nonumber\\
	&\bm G^{(S)}_{3}=\dfrac{i\omega}{4h_1}\Bigg([\bm p\times e_1\bm A_1 ]-[e_1\bm A_1\times\bm p]  \Bigg)\dfrac{1}{h_1}-\dfrac{i\omega}{4h_2}\Bigg([\bm p\times e_2\bm A_2 ]-[e_2\bm A_2\times\bm p]  \Bigg)\dfrac{1}{h_2}\,.
\end{align} 
A similar expression we have for the operator $\bm G^{(\Sigma)}$ in Eq.~\eqref{Hrad}, $\bm G^{(\Sigma)}=\bm G^{(\Sigma)}_{ 1}+\bm G^{(\Sigma)}_{2}+\bm G^{(\Sigma)}_{3}$, where
\begin{align}\label{Gsig}
	&\bm G^{(\Sigma)}_{1}= \dfrac{i}{2}\left\{\dfrac{1}{h_1},[\bm k\times e_1\bm A_1]\right\}-\dfrac{i}{2}\left\{\dfrac{1}{h_2},[\bm k\times e_2\bm A_2]\right\}\,,\nonumber\\
	&\bm G^{(\Sigma)}_{2}=  -\dfrac{1}{2h_1}\left(\dfrac{g}{r^3}-\dfrac{b}{r}\right)[\bm r\times e_1\bm A_1 ]\dfrac{1}{h_1}	-\dfrac{1}{2h_2}\left(\dfrac{g}{r^3}-\dfrac{b}{r}\right)[\bm r\times e_2\bm A_2 ]\dfrac{1}{h_2}\nonumber\\
	&+\dfrac{1}{2h_1}\dfrac{g}{r^3}[\bm r\times( e_1\bm A_1+e_2\bm A_2) ]\dfrac{1}{h_2}+\dfrac{1}{2h_2}\dfrac{g}{r^3}[\bm r\times( e_1\bm A_1+e_2\bm A_2) ]\dfrac{1}{h_1}\,,\nonumber\\
	&\bm G^{(\Sigma)}_{3}=\dfrac{i\omega}{4h_1}\Bigg([\bm p\times e_1\bm A_1 ]-[e_1\bm A_1\times\bm p]  \Bigg)\dfrac{1}{h_1}+\dfrac{i\omega}{4h_2}\Bigg([\bm p\times e_2\bm A_2 ]-[e_2\bm A_2\times\bm p]  \Bigg)\dfrac{1}{h_2}\,.
\end{align} 
Here $\omega=k$ is the photon energy.

Let us now proceed on to the calculation of matrix elements of radiative transitions.

\subsection{Transitions from  states with $L=1$ to  states with $L=0$}
All matrix elements of transitions from states with $L=1$ to states with $L=0$ are expressed in terms of matrix elements between states with the angular part of the wave function $Y_{1\mu}$ in the initial state and $Y_{00}$ in the final state.
Let $${\cal T}(J_2,\,\mu_2,\,s_2|J_1,\,\mu_1,\,s_1)$$ denotes the matrix element of the transition from a state with the total angular momentum $J_1$, projection $J_z=\mu_1$, total spin $S_{tot}=s_1$, and orbital angular momentum $L=1$ to a state with the total angular momentum $J_2$, projection $J_z=\mu_2$, total spin $S_{tot}=s_2=J_2$, and $L=0$. Then
\begin{align}\label{CalTtot}
	&{\cal T}(J_2,\,\mu_2,\,s_2|J_1,\,\mu_1,\,s_1)=\sum_{\mu'}C^{J_1\mu_1}_{1\mu',s_1 (\mu_1-\mu')}\langle s_2,\mu_2|\Bigg[(\bm e_{\mu'}\cdot\bm A_W^*)t^{(0)}\nonumber\\
	&+\dfrac{i}{2}\bm S_{tot}\cdot\Bigg([\bm e_{\mu'}\times\bm A_W^*]t_1^{(S)}+(\bm e_{\mu'}\cdot\bm n_k)[\bm n_k\times\bm A_W^*]t_2^{(S)}\Bigg )\nonumber\\
	&+\dfrac{i}{2}\bm \Sigma\cdot\Bigg([\bm e_{\mu'}\times\bm A_W^*]t_1^{(\Sigma)}+(\bm e_{\mu'}\cdot\bm n_k)[\bm n_k\times\bm A_W^*]t_2^{(\Sigma)}\Bigg )\Bigg]\,|s_1,\mu_1-\mu'\rangle\,,
\end{align} 
where $\bm n_k=\bm k/k$. All the matrix elements we need are expressed in terms of the coefficients $t^{(0)}$, $t^{(S)}_{1,2}$, and $t^{(\Sigma)}_{1,2}$. For transitions to the state with $J_2=1$, the matrix elements are
\begin{align}\label{CalTJ21}
	&{\cal T}(1,\,\mu,\,1|0,\,0,\,1)=\dfrac{1}{2\sqrt{3}}(\bm e_\mu^*\cdot \bm A_W^*)\left(-2t^{(0)}+2t_1^{(S)}+t_2^{(S)}\right)\,,\nonumber\\
	&{\cal T}(1,\,\mu_2,\,1|1,\,\mu_1,\,1)=\dfrac{i}{2\sqrt{2}}\Bigg[([\bm e_{\mu_2}^*\times\bm e_{\mu_1}]\cdot \bm A_W^*)\left(2t^{(0)}-t_1^{(S)}\right)\nonumber\\
	&+(\bm e_{\mu_2}^*\cdot\bm n_k)(\bm e_{\mu_1}\cdot[\bm n_k\times\bm A_W^*])t_2^{(S)}\Bigg]\,,\nonumber\\
	&{\cal T}(1,\,\mu_2,\,1|1,\,\mu_1,\,0)=\dfrac{i}{2}\Bigg[([\bm e_{\mu_2}^*\times\bm e_{\mu_1}]\cdot \bm A_W^*)t_1^{(\Sigma)}+(\bm e_{\mu_1}\cdot\bm n_k)(\bm e_{\mu_2}^*\cdot[\bm n_k\times\bm A_W^*])t_2^{(\Sigma)}\Bigg]	\,,\nonumber\\
	&{\cal T}(1,\,\mu_2,\,1|2,\,\mu_1,\,1)=(\bm e_{\mu_1-\mu_2}\cdot\bm A_W^*)\,C^{2\,\mu_1}_{1\,(\mu_1-\mu_2),1\mu_2}\left(t^{(0)}+\dfrac{t_1^{(S)}}{2}\right)+\dfrac{i}{2}t_2^{(S)}\nonumber\\
	&\times\sum_{\sigma=-1,0,+1}(\bm n_k\cdot\bm e_{\mu_1-\mu_2-\sigma})([\bm n_k\times\bm A_W^*]\cdot\bm e_\sigma)\,[\mu_2\delta_{\sigma,0}-\delta_{\sigma,1} +\delta_{\sigma,-1} ]C^{2\,\mu_1}_{1\,(\mu_1-\mu_2-\sigma),1(\mu_2+\sigma)}\,.
\end{align} 
Due to the selection rules, ${\cal T}(0,\,0,\,0|0,\,0,\,1)=0$. The remaining matrix elements of the transitions to the state with $J_2=0$ are
\begin{align}\label{CalTJ20}
	&{\cal T}(0,\,0,\,0|1,\,\mu,\,1)=\dfrac{1}{2\sqrt{2}}(\bm e_\mu\cdot \bm A_W^*)\left(2t_1^{(\Sigma)}+t_2^{(\Sigma)}\right)\,,\nonumber\\
	&{\cal T}(0,\,0,\,0|1,\,\mu,\,0)=(\bm e_{\mu}\cdot \bm A_W^*)t^{(0)}\,,\nonumber\\
	&{\cal T}(0,\,0,\,0|2,\,\mu,\,1)=\dfrac{i}{2}t_2^{(\Sigma)}\sum_{\sigma=-1,0,+1}(\bm n_k\cdot\bm e_{\mu-\sigma})([\bm n_k\times\bm A_W^*]\cdot\bm e_\sigma)\,C^{2\,\mu}_{1\,(\mu-\sigma),1\sigma}\,.
\end{align} 
The explicit expressions for the coefficients $t^{(0)}$, $t^{(S)}_{1,2}$, and $t^{(\Sigma)}_{1,2}$, calculated using the variational functions \eqref{WF0} and \eqref{WF1}, are presented in Appendix~\ref{appA}.
\subsection{Widths of radiative transitions }
Using the found amplitudes, we write the widths $\Gamma_{J_2,J_1}$ of radiative transitions  from states with $J_1$, $L=1$ to states with $J_2$, $L=0$. The width $\Gamma_{J_2,1}$ corresponds to the transition from a state with lower energy, and we denote the width of the transitions from a state with higher energy by $\Gamma'_{J_2,1}$, see Eq.~\eqref{psi12}. Further, in the amplitudes of radiative transitions, all charges of quarks and antiquarks are measured in units of the proton charge. Therefore, the widths are contain the fine structure constant $\alpha$ as a factor. For $J_2=1$ we have 
\begin{align}\label{Gam1}
	&\dfrac{\Gamma_{1,0}}{\Gamma_0}=\left(t^{(0)}-t_1^{(S)}-\dfrac{1}{2}t_2^{(S)}\right)^2\,,\nonumber\\
	&\dfrac{\Gamma_{1,1}}{\Gamma_0}=\left[ \left(t^{(0)}-\dfrac{1}{2}t_1^{(S)}\right)c_2+\dfrac{1}{\sqrt{2}}t_1^{(\Sigma)}c_1\right]^2+\dfrac{1}{8}\left(t_2^{(S)}c_2\right)^2+\dfrac{1}{4}\left(t_2^{(\Sigma)}c_1\right)^2 \,,\nonumber\\
 	&\dfrac{\Gamma'_{1,1}}{\Gamma_0}=\left[ \left(t^{(0)}-\dfrac{1}{2}t_1^{(S)}\right)c_1-\dfrac{1}{\sqrt{2}}t_1^{(\Sigma)}c_2\right]^2+\dfrac{1}{8}\left(t_2^{(S)}c_1\right)^2+\dfrac{1}{4}\left(t_2^{(\Sigma)}c_2\right)^2 \,,\nonumber\\
 	&\dfrac{\Gamma_{1,2}}{\Gamma_0}= \left(t^{(0)}+\dfrac{1}{2}t_1^{(S)}+\dfrac{1}{4}t_2^{(S)}\right)^2+\dfrac{9}{80}(t_2^{(S)})^2\,,
\end{align} 
where $\Gamma_0=4\alpha k/3$. 

For $J_2=0$ the widths are
\begin{align}\label{Gam0}
	&\dfrac{\Gamma_{0,1}}{\Gamma_0}=\left[ t^{(0)}c_1+\dfrac{1}{2\sqrt{2}}\left(2t_1^{(\Sigma)}+t_2^{(\Sigma)}\right)c_2\right]^2\,,\nonumber\\
	&\dfrac{\Gamma'_{0,1}}{\Gamma_0}=\left[ t^{(0)}c_2-\dfrac{1}{2\sqrt{2}}\left(2t_1^{(\Sigma)}+t_2^{(\Sigma)}\right)c_1\right]^2\,,\nonumber\\
	&\dfrac{\Gamma_{0,2}}{\Gamma_0}=\dfrac{3}{40}\left(t_2^{(\Sigma)}\right)^2	\,.
\end{align} 
We recall that $\Gamma_{0,0}=0$ by virtue of the selection rules.

A comparison of predictions, obtained withing the relativistic potential model, with experiment for the partial width $\Gamma^*$ of the radiative transition from the state with $L=0$ and $J=1$ to the state with $L=0$ and $J=0$ is performed in detail in Ref.~\cite{BM2025b}. This width has the form
\begin{align}\label{Gamstar}
	&\dfrac{\Gamma^*}{\Gamma_0}=[e_1\bar\tau_0-e_2\tau_0]^2\,.
\end{align}
For the reader's convenience, we provide the functions $\bar\tau_0$ and  $\tau_0$ in Appendix~\ref{appB}. 

Calculating the value of photon energy $k$, the recoil during emission must be taken into account. This value  is related to the energy difference $\nu_{fi}=M_i-M_f$ between the levels as follows $$k=\dfrac{M_i^2-M_f^2}{2M_i}=\nu_{fi}-\dfrac{\nu_{fi}^2}{2M_i}\,,$$
where $M_i$ and $M_f$ are the energies of the initial and final levels.

Of interest are also radiative transitions from states with $J=1$ and $L=1$ to state with $J=0$ and $L=1$, since the partial widths of these transitions depend strongly on both the mixing coefficients and the relativistic corrections. The amplitudes  $T_1$ and $T_1'$ of the transitions from the states $\Psi_1$ and $\Psi_1'$, see Eq.~\eqref{psi12}, have the form
\begin{align}\label{T1T0}
	&T_1=i(\bm e_\mu\cdot[\bm n_k\times\bm A^*_W])\left[c_1 T^{(\Sigma)}+c_2\left(T^{(0)}+T^{(S)}\right)\right]\,,\nonumber\\
	&T_1'=i(\bm e_\mu\cdot[\bm n_k\times\bm A^*_W])\left[c_2 T^{(\Sigma)}-c_1\left(T^{(0)}+T^{(S)}\right)\right]\,,
\end{align}
where $J_z=\mu$ in the initial state. The corresponding partial widths $\Gamma_1$ and $\Gamma_1'$ are 
\begin{align}\label{Gam11Gam01}
	&\dfrac{\Gamma_{1}}{\Gamma_0}=\left[ c_1T^{(\Sigma)}+c_2\left(T^{(0)}+T^{(S)}\right)\right]^2 \,,\nonumber\\
	&\dfrac{\Gamma_{1}'}{\Gamma_0}=\left[ c_2T^{(\Sigma)}-c_1\left(T^{(0)}+T^{(S)}\right)\right]^2 \,.
\end{align} 
The explicit expressions for the coefficients $T^{(0)}$, $T^{(S)}$
and $T^{(\Sigma)}$, as well as our predictions, are given in Appendix~\ref{appC}.

\section{Predictions and comparison with experiment}
To obtain numerical predictions for the radiative transition amplitudes, we must first fix the values of the model  parameters. The choice of the parameters is based to a large extent on the use of known experimental data for the meson mass spectrum.
Unfortunately, experimental data for partial widths of radiative transitions are currently very limited. Therefore, the choice of parameters in our model has some degree of freedom. We have checked  that a variation of the parameters that maintains qualitative agreement with the available experimental data for the transition widths leads to a change in the predictions  of order  30\%.
Also, some uncertainty in the accuracy of the predictions arises from the use of the variational method to determine the wave functions. We have found that this uncertainty is significantly smaller than that related to the limited amount of available experimental data. It turned out that taking into account the indicated uncertainties does not significantly change the predictions for the mass spectra and the hierarchy of partial radiative transition widths.

 We choose the parameter $b$ in the Hamiltonian \eqref{H0} to be $b=0.1\,\mbox{GeV}^2$. This value has been obtained in Ref.~\cite{BM2025b} by a detailed analysis of the hyperfine splitting between the energy levels with $L=0,\,S=1$ and $L=0,\,S=0$ in systems with different heavy and light quarks. 

\subsection{Systems $b\bar b$ and $c\bar c$}\label{bbcc}
Having fixed the value of $b$, we determined the masses of the $c$ and $b$ quarks by analyzing the mass spectrum in the $c\bar c$ and $b\bar b$ systems. They turned out to be $m_c=1.6\, \mbox{GeV}$ and $m_b=5\,\mbox{GeV}$. For different systems, the values of the constant $g$ are different, since they correspond to the average value of the running coupling constant of the strong interaction over the wave functions.
When a characteristic size of the wave function decreases, the value of $g$ becomes smaller. For the $c\bar c$ and $b\bar b$ systems, these values are $g=0.78$ and $g=0.57$, respectively.

The variational parameters $\omega_{0,1}$ in the wave functions turned out to be $\omega_0=0.709\,\mbox{GeV},\,\omega_1=0.465\,\mbox{GeV} $ for $c\bar c$  and $\omega_0=1.277\,\mbox{GeV},\,\omega_1=0.734\,\mbox{GeV} $ for $b\bar b$. Moreover, the values of the constant ${\cal C}$ in the Hamiltonian \eqref{H0}, in general, depend on $L$, that is, ${\cal C}\equiv {\cal C}_L$. These values do not affect the level splitting and the matrix elements of radiative transitions, but only the overall shift of levels in a given multiplet. We have ${\cal C}_0=-0.022\,\mbox{GeV} $, ${\cal C}_1=0.043\,\mbox{GeV}$ for $c\bar c$ and  ${\cal C}_0=-0.264\,\mbox{GeV} $, ${\cal C}_1=-0.234\,\mbox{GeV}$ for $b\bar b$. A comparison of the model predictions for the mass spectrum with experimental data for these systems is performed in Table~\ref{mccbb}.
 \begin{table}[!h]
	\caption{\label{mccbb} Prediction for the mass spectrum of $ c \bar c$ and $ b \bar b$ systems (in \mbox{MeV}) and comparison with experiment.}
	\begin{center}
		\begin{tabular}{lcccccc} 
			&$M(\eta_c)$ &$M(J/\psi)$ &$M(\chi_{c0})$ &$M(\chi_{c1})$ &$M (h_{c})$ &$M (\chi_{c2})$\\
			\hline
			
			Predictions &2984 &3093 &3466 &3513 &3525 &3544\\
			
			Experiment \cite{Navas:2024ynf} &2984 &3097 &3415 &3511 &3525 &3556\\
			\hline
			
			&$M(\eta_b)$  &$M(\Upsilon)$ &$M(\chi_{b0})$  &$M(\chi_{b1})$ &$M(h_{b})$ &$M(\chi_{b2})$\\
			\hline
			
			Predictions &9399 &9459 &9876 &9894 &9899 &9907\\
			
			Experiment \cite{Navas:2024ynf} &9399 &9460 &9859 &9893 &9899 &9912\\
			\hline
		\end{tabular}
	\end{center}
\end{table}
Note that for $c\bar c$ and $b\bar b$ the total spin $S_{tot}$ is conserved. In this case, the solutions with $L=1,\,S_{tot}=1$ correspond to $\chi_{c1}$ and $\chi_{b1}$, and the solutions with $L=1,\,S_{tot}=0$ correspond to $h_{c}$ and $h_{b}$. It is also interesting to compare the predictions of our model for the widths of radiative transitions in the $c\bar c$ and $b\bar b$ systems with the available experimental data and with predictions of some models, see Table~\ref{ccgam}. It is seen  that the predictions of our model is in reasonable agreement with available experimental data for charmonium. There are no measured radiation widths for bottomonium, but comparison with other calculations shows agreement at a $30\%$ accuracy level.
\begin{table}[!h]
	\caption{\label{ccgam} Predictions of the partial widths of radiative transitions (in \mbox{keV}) for the $c\bar c$ and $b\bar b$ systems in comparison with the predictions of some models that take into account relativistic corrections to the widths of radiative transitions.}
	\begin{center}
		\begin{tabular}{lccccc} 
			&$\Gamma(\chi_{c0}\to J/\psi\gamma)$  &$\Gamma(\chi_{c1}\to J/\psi\gamma)$ &$\Gamma(\chi_{c2}\to J/\psi\gamma)$  &$\Gamma(J/\psi\to \eta_c\gamma)$ &$\Gamma(h_c\to \eta_c\gamma)$\\
			\hline
			
			This paper &136 &142  &158 &1.51  &190\\
			 
            Relativistic QM \cite{Ebert:2002pp} &121 &265  &327 &1.05 & 560\\ 

            Relativized GI \cite{Barnes:2005pb} &114 &239  &313 &2.4 &352\\ 

            \cite{Radford:2007vd} &142.2 &287.0  &390.6 &2.7 &610.0\\ 

            Lattice QCD \cite{Dudek:2009kk} &199(6) &270(70) &380(50) &2.51(8) &\dots\\ 

            Screened Potential \cite{Li:2009zu} &117 &244 &309 &\dots &323\\ 
            
			Experiment \cite{Navas:2024ynf} &$148\pm 15$ &$292\pm 22$ &$360\pm 40$ &$1.31\pm 0.13$ &$470\pm 170$\\ 
			\hline
			
			&$\Gamma(\chi_{b0}\to \Upsilon\gamma)$  &$\Gamma(\chi_{b1}\to \Upsilon\gamma)$ &$\Gamma(\chi_{b2}\to \Upsilon\gamma)$  &$\Gamma(\Upsilon\to \eta_b\gamma)$ &$\Gamma(h_b\to \eta_b\gamma)$\\
			\hline   
			This paper &19.42  &19.96 &21.63 &0.008  &23.76\\ 
            
			Relativistic QM \cite{Ebert:2002pp} &29.9 &36.6 &40.2 &0.0058 &52.6\\ 

            Relativized GI \cite{Godfrey:2015dia} &23.8 &29.5  &32.8 &0.010 &35.7\\ 

            \cite{Radford:2007vd} &22.1 &27.3  &31.2 &0.004 &37.9\\ 

             Screened Potential \cite{Li:2009nr} &24.3 &30.0 &32.6 &\dots &36.3\\ 
             
			\hline
		\end{tabular}
	\end{center}
\end{table}

\subsection{Systems $b\bar u$ and $b\bar d$}\label{qb}
The masses of charged meson $b\bar u$ and  neutral meson $b\bar d$ are practically the same. In this and the next sections, we will assume that the masses of quarks $m_u$ and $m_d$ are $m_q=(m_u+m_d)/2$. The constants $g$ and $m_q$ that provide an agreement between the model for the mass spectrum and experimental data are $m_q=0.25\,\mbox{GeV}$ and $g=0.94$. For this parameter, we have $\omega_0=0.715\,\mbox{GeV}$, $\omega_1=0.412\,\mbox{GeV}$, ${\cal C}_0=0.077\,\mbox{GeV}$ and ${\cal C}_1=0.115\,\mbox{GeV}$. The corresponding prediction for the mass spectrum and its comparison with the experimental values are given in Table~\ref{mqb}.
\begin{table}[!h]
	\caption{\label{mqb} Prediction for the mass spectrum of the $b\bar u$ and $b\bar d $ systems (in \mbox{MeV}) and comparison with experiment.}
	\begin{center}
		\begin{tabular}{lcccccc} 
			&$M(B^{(-)})$  &$M(B^{{(-)}*})$ &$M(B^{(-)}_{0})$  &$M(B^{(-)}_{1})$ &$ M (B^{(-)}_{1}{}')$  &$ M (B^{(-)}_{2})$\\
			\hline
			
			Predictions &5280 &5382 &5677 &5711 &5726 &5740\\
			
			Experiment \cite{Navas:2024ynf} &5279 &5325 &\dots &\dots &5726 &5737\\
			\hline
			
			&$M(B^{(0)})$  &$M(B^{(0)*})$ &$M(B^{(0)}_{0})$ &$M(B^{(0)}_{1})$ &$ M (B^{(0)}_{1}{}')$  &$ M (B^{(0)}_{2})$\\
			\hline
			
			Predictions &5280 &5382 &5677 &5711 &5726 &5740\\
			
			Experiment \cite{Navas:2024ynf} &5280 &5325 &\dots &\dots &5726 &5740\\
			\hline
		\end{tabular}
	\end{center}
\end{table}
It is seen  good agreement between  predictions and experimental data.

The constants $c_1$, $c_2$ and the mixing angle $\varphi$ are $c_1=-0.382$, $c_2=0.924$, and $\varphi=-22.46^\circ$. The quantity $W_{1/2}$ \eqref{DHSinf3} is $0.05$. It is clear that the mixing angle prediction for the $b\bar u$  and $b \bar d$ systems is close to that obtained in the infinite-heavy quark theory. The partial widths of the radiative transitions $b\bar u$ and $b\bar d$ are given in Table~\ref{tb_bd} and Table~\ref{tb_bu}.
\begin{table}[!h]
	\caption{\label{tb_bd} Predictions for radiative decay widths  $B_{J}^{(0)}\to B^{(0)}\gamma$ and $B_{J}^{(0)}\to B^{(0)*}\gamma$ in \mbox{keV}.}
	\begin{center}
		\begin{tabular}{lccccc} 
			Publication &\cite{li:2021hss} &\cite{Godfrey:2016nwn} &\cite{Lu:2016bbk} &\cite{Asghar:2018tha} &This work\\
			\hline
			
			$\Gamma(B_{0}^{(0)}\to B^{(0)*}\gamma)$ &149 &92.7 &116.9 &175 &36\\
			
			$\Gamma(B_{1}^{(0)}\to B^{(0)}\gamma)$ &38 &106 &130.2 &137 &25.7\\
			
			$\Gamma(B_{1}^{(0)}\to B^{(0)*}\gamma)$ &24 &27.8 &53.1 &103 &28.2\\
			
			$\Gamma(B_{1}^{(0)}{}'\to B^{(0)}\gamma)$ &24 &37.8 &60.4 &127 &15\\
			 
			$\Gamma(B_{1}^{(0)}{}'\to B^{(0)*}\gamma)$ &66 &85.5 &108.5 &137 &2.9\\
			
			$\Gamma(B_{2}^{(0)}\to B^{(0)*}\gamma)$ &51 &126 &177.7 &232 &17.6\\
			\hline
		\end{tabular}
	\end{center}
\end{table}
\begin{table}[!h]
	\caption{\label{tb_bu} Predictions for radiative decay widths  $B_{J}^{(-)}\to B^{(-)}\gamma$  and $B_{J}^{(-)}\to B^{(-)*}\gamma$ in \mbox{keV}.}
	\begin{center}
		\begin{tabular}{lcccc} 
			Publication &\cite{li:2021hss} &\cite{Godfrey:2016nwn} &\cite{Asghar:2018tha} &This work\\
			\hline
			
			$\Gamma(B_{0}^{(-)}\to B^{(-)*}\gamma)$ &477 &325 &575 &137.5\\
			
			$\Gamma(B_{1}^{(-)}\to B^{(-)}\gamma)$ &111 &373 &448 &111.2\\
			
			$\Gamma(B_{1}^{(-)}\to B^{(-)*}\gamma)$  &75 &97.5 &339 &101.4\\
			
			$\Gamma(B_{1}^{(-)}{}'\to B^{(-)}\gamma)$  &69 &132 &415 &28.8\\
			
			$\Gamma(B_{1}^{(-)}{}'\to B^{(-)*}\gamma)$  &206 &300 &448 &7.7\\
			
			$\Gamma(B_{2}^{(-)}\to B^{(-)*}\gamma)$ &146 &444 &761 &35.8\\
			\hline		
		\end{tabular}
	\end{center}
\end{table}

At the moment, unfortunately, there are no experimental data on the radiative widths of the P-wave states of B mesons, so there is nothing left to do but compare our results with other calculations. From Table~\ref{tb_bd} and Table~\ref{tb_bu} it is clear that our model predicts significantly smaller values for decay widths of $B_{1}'\to B^*\gamma$ for both charged and neutral B mesons as compared with predictions of other models. 

\subsection{System $b\bar s$}
For the system $b\bar s$, using  the above values of $b$ and $m_b$,  we find $m_s=0.55\,\mbox{GeV}$, $g=0.82$, $\omega_0=0.715\,\mbox{GeV}$, $\omega_1=0.442\,\mbox{GeV}$, ${\cal C}_0=-0.098\,\mbox{GeV}$ and ${\cal C}_1=-0.046\,\mbox{GeV}$. The coefficients $c_1$, $c_2$ and the corresponding mixing angle $\varphi$ are $c_1=-0.392$, $c_2=0.92$ and $\varphi=-23.1^\circ$. The quantity $W_{1/2}$ \eqref{DHSinf3} is $0.04$ which is also close to the prediction of  infinite-heavy quark theory. For these parameters, our predictions for the mass spectrum and partial radiative widths are given in Table~\ref{mbs} and Table~\ref{tb_bs}.
\begin{table}[!h]
	\caption{\label{mbs} Prediction for the mass spectrum of the $b\bar s $ system (in \mbox{MeV}) and comparison with experiment.}
	\begin{center}
		\begin{tabular}{lcccccc} 
			&$M(B_s)$ &$M(B_s^{*})$ &$M(B_{s0})$ &$M(B_{s1})$ &$M(B'_{s1})$ &$M(B_{s2})$\\
			\hline
			
			Predictions &5367 &5443 &5786 &5815 &5828 &5840\\
			
			Experiment \cite{Navas:2024ynf} &5367 &5415 &\dots &\dots &5829 &5840\\
			\hline
		\end{tabular}
	\end{center}
\end{table}
\begin{table}[!h]
	\caption{\label{tb_bs} Predictions for radiative decay widths  $B_{sJ}\to B_s\gamma$ and $B_{sJ}\to B_s^*\gamma$  in \mbox{keV}.}
	\begin{center}
		\begin{tabular}{lccccc} 
			Publication &\cite{li:2021hss} &\cite{Godfrey:2016nwn} &\cite{Lu:2016bbk} &\cite{Asghar:2018tha} &This work\\
			\hline
			
			$\Gamma(B_{s0}\to B_s^*\gamma)$ &102 &76 &84.7 &133 &46.8\\
			
			$\Gamma(B_{s1}\to B_s\gamma)$ &37 &70.6 &97.7 &\dots &19.9\\
			
			$\Gamma(B_{s1}\to B_s^*\gamma)$ &56 &36.9 &39.5 &\dots &35\\
			
			$\Gamma(B_{s1}'\to B_s\gamma)$ &27 &47.8 &56.6 &\dots &24.2\\
			 
			$\Gamma(B_{s1}'\to B_s^*\gamma)$ &53 &57.3 &98.8 &\dots &4.1\\
			
			$\Gamma(B_{s2}\to B_s^*\gamma)$ &51 &106 &159 &225 &27.2\\
			\hline
		\end{tabular}
	\end{center}
\end{table}

It should be noted here that the model predicts the masses of $B_{s0}$ and $B_{s1}$ above the threshold for $BK$ and $B^*K$, respectively. If this is indeed the case, then these states are expected to be very wide. However, if, as in the case of $D_{s0}$ and $D_{s1}$ (see below), due to possible mixing with the molecular states, the masses of these particles become smaller than the threshold values. As a result,  their widths will be small and the radiative decays will likely be dominant. Therefore, we believe that searching for these states in radiative decays in the Belle II and LHC experiments appears to be very promising.

As for the radiative decay of $B_{s2}$, the probability of its decay is quite small, according to our model, on the order of $2\%$. And therefore it is unlikely to be observed in the near future.

The conclusion on the possibility to observe the radiative decay of $B_{s1}'$ is not so clear-cut. The full width of $B_{s1}'$ is poorly known at present. Its value, according to PDG \cite{Navas:2024ynf}, equals $0.5\pm 0.3\pm 0.3\,\mbox{MeV}$, i.e., it is consistent with the zero value. If we estimate the partial width of the dominant decay of this state into $D^*K$ from the similar decay of $D_{s2} $ in the limit of a heavy $b$ quark, for example, following Ref.~\cite{Colangelo:2012xi}, then it turns out to be of the order of $17\pm 5\,\mbox{keV}$, which may mean that the probability of radiative decays is significant and, possibly, greater than the strong decay to $D^*K$. Of course, it should be noted that we do not know the exact mixing angle with the state $b\bar s$ ($j =1/2$), the admixture of which can radically increase the total width of $B_{s1}'$. Nevertheless, it should be noted that in our model, unlike previously published ones, the main radiative decay is $B_{s1}'\to B_s\gamma$, so the search for such a decay seems quite promising to us.

\subsection{System $b\bar c$}
The $b\bar c$ system is interesting because both particles (quark and antiquark) are heavy, but the antiquark $\bar c$ is not heavy enough for QCD to be used with high accuracy in the $b\bar c$ system. Although a considerable number of papers have been devoted to the theoretical study of this system (see Ref.~\cite{Akan:2025nej} and the literature cited therein), the experimental data are very limited \cite{Navas:2024ynf}. Since the reduced mass in the $b\bar c$ system is smaller than the reduced mass in $b\bar b$ and larger than in $b\bar s$, it is natural to expect that $g(b\bar b)<g(b\bar c)<g(b\bar s)$. Therefore, for our estimates, we chose $g(b\bar c)=0.75$. After this, we obtained the value ${\cal C}_0=-0.022\,\mbox{GeV}$ from a comparison of the model prediction with the experiment for the state $B_c$. It can also be assumed that the value ${\cal C}_1$ for $b\bar c$ is about the corresponding value for $b\bar s$. For the estimate, we chose it  to be ${\cal C}_1=-0.034\,\mbox{GeV}$. The corresponding values for $\omega_{0,1}$ are $\omega_{0}=0.982\,\mbox{GeV}$ and $\omega_{1}=0.57\,\mbox{GeV}$. The coefficients $c_1$, $c_2$ and the corresponding mixing angle $\varphi$ are $c_1=-0.373$, $c_2=0.928$, and $\varphi=-21.9^\circ$, the corresponding value of $W_{1/2}$ is $0.05$. The result of our predictions for the mass spectrum is given in the Table~\ref{mbc}.
\begin{table}[!h]
	\caption{\label{mbc} Prediction for the mass spectrum of the $b\bar c$ system (in \mbox{MeV}) and comparison with experiment.}
	\begin{center}
		\begin{tabular}{lcccccc} 
			&$M(B_c)$ &$M(B_c^{*})$ &$M(B_{c0})$  &$M(B_{c1})$ &$ M(B'_{c1})$  &$M(B_{c2})$\\
			\hline
			
			Predictions &6275 &6367 &6727 &6756 &6769 &6782\\
			
			Experiment \cite{Navas:2024ynf} &6275 &\dots &\dots &\dots &\dots &\dots\\
			\hline
		\end{tabular}
	\end{center}
\end{table}
Our predictions for the mass spectrum in the $b\bar c$ system are in qualitative agreement with the predictions of other authors \cite{Akan:2025nej,Soni:2017wvy,Mutuk:2018erw,Devlani:2014nda,Ebert:2011jc,Godfrey:2004ya,Eichten:1994gt,Monteiro:2016rzi}. The predictions of our model for the photon energies $k$ and partial widths of radiative transitions in the $b\bar c$ system are given in Table~\ref{tb_bc}.

In previous works, only the electric dipole moment operator in the nonrelativistic approximation was used to calculate the $E1$ transitions, just as the nonrelativistic magnetic moment operator was used to calculate the magnetic $M1$ transition amplitudes, and the remaining corrections, in contrast to our work, were not taken into account \cite{Monteiro:2016rzi,Godfrey:2004ya,Eichten:1994gt,Devlani:2014nda,Soni:2017wvy}, which led to visible difference in the transition widths.
\begin{table}[!h]
	\caption{\label{tb_bc} The photon energies $k$ (in \mbox{MeV}) and widths (in \mbox{keV}) of the El electromagnetic transitions in the $b\bar c$ system.}
	\begin{center}
		\begin{tabular}{lcccccccc} 
			Transitions &\cite{Gershtein:1994dxw} &\cite{Li:2023wgq} &\cite{Li:2022bre} &\cite{Li:2019tbn} &\cite{Eichten:2019gig} &\cite{Godfrey:2004ya} &\cite{Ebert:2002pp} &Ours\\
			\hline

			\multirow{2}{*}{$B_{c0}\to B_{c}^*\gamma$} &366 &353 &379 &377 &354 &358 &355  &351\\
			&65.3 &52.3 &59.58 & 96 &53.1 &55 &67.2  &51.1\\
			
			\multirow{2}{*}{$B_{c1}\to B_{c}^*\gamma$} &400 &395 &433 &416 &389 &391 &389 &378\\
			&77.8 &47.8 &71.52 &70 &62.5 &60 &78.9 & 45.6\\
			
			\multirow{2}{*}{$B_{c1}\to B_{c}\gamma$} &464 &457 &478 &468 &440 &454 &447 &464\\
			&11.6 &30.1 &16.5 &35 &9.9 &13 &18.4 &21.5\\
			
			\multirow{2}{*}{$B_{c1}'\to B_{c}^*\gamma$} &412 &403 &425 &433 &397 &399 &405 &391\\
			&8.10 &25.6 &13.75 &40 &7.5 &11 &13.6 &3.3\\
			
			\multirow{2}{*}{$B_{c1}'\to B_{c}\gamma$} &476 &466 &470 &484 &448 &462 &463 &477\\ 
			&131. &64.0 &76.5 &74 &92.4 &80 &132 &37.9\\
			
			\multirow{2}{*}{$B_{c2}\to B_{c}^*\gamma$} &426 &421 &445 &445 &409 &416 &416 &402\\
			&102.9 &85.1 &90 &87 &79.7 &83 &107 &43.2\\
			\hline
		\end{tabular}
	\end{center}
\end{table}

Recently, the LHCb experiment reported the discovery of what appear to be radiative decays of the P-wave states of $B_c$ mesons \cite{LHCb:2025uce,LHCb:2025ubr}. Since the masses of these states are significantly smaller than the decay threshold to $B^{(*)}D$, it can be expected that the dominant decays of all four P-wave states will be radiative decays. Due to the limited energy resolution of the LHCb detector for gamma quanta, the six possible radiative transition peaks overlap significantly. The energies and partial widths of the radiative decays obtained in our model are in reasonable agreement with the observed experimental picture. It is hoped that in the future, using events with the conversion of gamma quanta into $e^+e^-$ pairs, it will be possible to observe the individual radiative transitions and to measure the relative probabilities of $B_{c1}\to B_c\gamma$, $B_{c1}\to B_c^*\gamma$, and $B_{c1}'\to B_c\gamma$, $B_{c1}'\to B_c^*\gamma$ decays.

\subsection{Systems $c\bar u$ and $c\bar d$}\label{qc}
For the neutral  $c\bar u $ and charged $c\bar d$ mesons, the masses of the light antiquarks are the same as in the $b\bar u $ and $b\bar d$ mesons, and the constant $g$ providing the agreement between the model for the mass spectrum and the experimental data is $g=0.95$.
 For these parameters we have $\omega_0=0.528\,\mbox{GeV}$, $\omega_1=0.359\,\mbox{GeV}$, ${\cal C}_0=0.095\,\mbox{GeV}$ and ${\cal C}_1=0.209\,\mbox{GeV}$. The corresponding 
predictions for the mass spectrum and their comparison with experimental values
are given in Table~\ref{mqc}. 
\begin{table}[!h]
	\caption{\label{mqc} Prediction for the mass spectrum of $c\bar u $ and $c\bar d $ systems (in \mbox{MeV}) and comparison with experiment.}
	\begin{center}
		\begin{tabular}{lcccccc} 
			&$M(D^{(0)})$ &$M(D^{(0)*})$ &$M(D^{(0)}_{0})$ &$M(D^{(0)}_{1})$ &$M(D^{(0)}_{1}{}')$ &$M(D^{(0)}_{2})$\\
			\hline
			
			Predictions &1867 &2020 &2343 &2413 &2427 &2452\\
			
			Experiment \cite{Navas:2024ynf} &1865 &2007 &2343 &2412 &2422 &2461\\
			\hline
			
			&$M(D^{(+)})$ &$M(D^{(+)*})$ &$M(D^{(+)}_{0})$ &$M(D^{(+)}_{1})$ &$M(D^{(+)}_{1}{}')$ &$M(D^{(+)}_{2})$\\
			\hline
			
			Predictions &1867 &2020 &2343 &2413 &2427 &2452\\
			
			Experiment \cite{Navas:2024ynf} &1870 &2010 &2343 &\dots &2426 &2464\\
			\hline
		\end{tabular}
	\end{center}
\end{table}
Good agreement is evident between  predictions and the experimental data.

The constants $c_1$, $c_2$, and the mixing angle $\varphi$ are $c_1=0.149$, $c_2=0.989$, and $\varphi=8.57^\circ$. The corresponding value of $W_{1/2}$ is $0.48$. Naturally, the partial widths of the radiative transitions $c\bar u$ and $c\bar d$ differ greatly due to the different charges of the antiquarks, see Table~\ref{tb_cu_cd}.
\begin{table}[!h]
	\caption{\label{tb_cu_cd} Prediction for partial radiative transition widths (in \mbox{keV}) in the $c\bar u$ and $c\bar d$ systems.}
	\begin{center}
		\begin{tabular}{lcccccc} 
			Publication &\cite{Li:2022vby} &\cite{Godfrey:2015dva} &\cite{Close:2005se} &\cite{Colangelo:1993zq} &\cite{Korner:1992pz} &This work\\
			\hline
			
			$\Gamma(D_{0}^{(-)}\to D^{(-)*}\gamma)$ &28.5 &30 &17 &$<2.8$ &\dots &18.3\\
			
			$\Gamma(D_{1}^{(-)}\to D^{(-)}\gamma)$ &26.4 &66 &19.7 &\dots  &\dots &104.2\\
			
			$\Gamma(D_{1}^{(-)}\to D^{(-)*}\gamma)$ &30.0 &8.6 &\dots &\dots &\dots &3.4\\
			
			$\Gamma(D_{1}^{(-)}{}'\to D^{(-)}\gamma)$ &55.1 &16.1 &\dots &$<3.3$ &\dots &1.6\\ 
			
			$\Gamma(D_{1}^{(-)}{}'\to D^{(-)*}\gamma)$ &16.1 &39.9 &30.9 &$<2.3$ &\dots &36.7\\
			
			$\Gamma(D_{2}^{(-)}\to D^{(-)*}\gamma)$  &59.7 &61.2 &51 &\dots &\dots &15.4\\
			\hline
			
            $\Gamma(D_{0}^{(0)}\to D^{(0)*}\gamma)$ &275.4 &288 &304 &$115\pm 54$ &\dots &101.6\\
            
			$\Gamma(D_{1}^{(0)}\to D^{(0)}\gamma)$ &254.9 &640 &349.3 &\dots &$245\pm 18$ &94.6\\
			
			$\Gamma(D_{1}^{(0)}\to D^{(0)*}\gamma)$ &290.2 &82.8 &\dots &\dots &$60\pm 5$ &68\\
			
			$\Gamma(D_{1}^{(0)}{}'\to D^{(0)}\gamma)$ &533.1 &156 &\dots &$14\pm 6$ &\dots &103.5\\
			 
			$\Gamma(D_{1}^{(0)}{}'\to D^{(0)*}\gamma)$ &155.4 &386 &549.5 &$93\pm 44$ &\dots &45.1\\
			
			$\Gamma(D_{2}^{(0)}\to D^{(0)*}\gamma)$  &577.4 &592 &895 &\dots &\dots &34.5\\
			\hline
		\end{tabular}
	\end{center}
\end{table}

\subsection{System $c\bar s$}
The $c\bar s$ system is the most complicated for analyzing radiative transitions. All numerous attempts to explain the hierarchy of partial widths of radiative transitions based on the use of nonrelativistic electric dipole and magnetic moments have proven unsuccessful. A detailed review of the works containing the corresponding predictions for electric dipole transitions \cite{Godfrey:2005ww,Goity:2000dk, Green:2016occ,Radford:2009bs,Chen:2020jku,Korner:1992pz} and magnetic dipole transitions \cite{Tran:2023hrn,Pullin:2021ebn,Becirevic:2009xp,Donald:2013sra,Deng:2013uca,Colangelo:1993zq,Ebert:2002xz,Choi:2007se,Zhu:1996qy,Aliev:1994nq,Close:2005se} is presented in Refs.~\cite{BM2025a, BM2025b}. 
 
We found that good agreement between the model prediction for energy levels and the experimental data is achieved at $g=0.93$. In this case, the values of $\omega_0$ and $\omega_1$ in the variational wave functions are $\omega_0=0.594\,\mbox{GeV}$  and $\omega_1=0.393\,\mbox{GeV}$. Fulfillment of the ratio $\omega_0>\omega_1$ for all systems is due to the fact that a characteristic size of the wave function with the lowest energy for $L=1$ is larger than this size for the ground state  with $L=0$. For the parameters ${\cal C}_{0,1}$ in the Hamiltonian $H^{(0)}$ in Eq.~\eqref{H0} we have ${\cal C}_0=-0.006\,\mbox{GeV}$ and ${\cal C}_1=0.07\,\mbox{GeV}$. The model predictions for energy levels (particle masses) are shown in Table~\ref{mcs}.
\begin{table}[!h]
	\caption{\label{mcs} Prediction for the mass spectrum of the $c\bar s $ system (in \mbox{MeV}) and comparison with experiment.}
	\begin{center}
		\begin{tabular}{lcccccc} 
			&$M(D_s)$  &$M(D_s^{*})$ &$M(D_{s0})$ &$M(D_{s1})$ &$M(D'_{s1})$ &$M(D_{s2})$\\
			\hline
			
			Predictions &1968 &2122 &2455 &2525 &2542 &2569\\
					
			Experiment \cite{Navas:2024ynf} &1968 &2112 &2317 &2460 &2535 &2569\\
			\hline
		\end{tabular}
	\end{center}
\end{table}

For our parameters we obtain  $c_1=-0.117$, $c_2=0.993$, $\varphi=-6.72^\circ$, and $W_{1/2}=0.23$. It is clear that the infinitely heavy quark approximation for the $c\bar s $ system is not satisfactory. The model predictions for the partial radiative transition widths are given in Table~\ref{tb_cs}.
\begin{table}[!h]
	\caption{\label{tb_cs} Predictions for radiative decay widths $D_{sJ}\to D_s\gamma$ and $D_{sJ}\to D_s^{*}\gamma$ in \mbox{keV}.}
	\begin{center}
		\begin{tabular}{lccccccc} 
			Publication &\cite{Godfrey:2005ww} &\cite{Goity:2000dk} &\cite{Green:2016occ} &\cite{Radford:2009bs} &\cite{Chen:2020jku} &\cite{Korner:1992pz} &This work\\
			\hline
				
			$\Gamma(D_{s0}\to D_s^{*}\gamma)$ &1.9 &14.5-24.9 &5.46 &4.92 &2.06-2.07 &\dots &4.51\\
				
            $\Gamma(D_{s1}\to D_s\gamma)$ &6.2 &10.3-17.2 &13.2 &12.8 &3.53-3.61 &\dots &49.66\\
                 
			$\Gamma(D_{s1}\to D_s^{*}\gamma)$ &5.5 &14.0-25.1 &17.4 &15.5 &4.74-4.79 &\dots &1.95\\
				
			$\Gamma(D_{s1}'\to D_s\gamma)$ &15 &25.2-31.1 &61.2 &54.5 &18.18-18.85 &$1.6\pm 2.3$ &1.51\\
				
			$\Gamma(D_{s1}'\to D_s^{*}\gamma)$ &5.6 &14.6-22.8 &9.21 &8.9 &2.96-3.02 &$10.4\pm 1.$ &16.35\\
				
            $\Gamma(D_{s2}\to D_s^{*}\gamma)$ &19 &41.5-55.9 &49.6 &44.1 &15.23-15.66 &$9.4\pm 2.0$ &3.75\\
			\hline
		\end{tabular}
	\end{center}
\end{table}	

It is seen that the mixing angles for $c\bar s$,  $c\bar u$, and $c\bar d$ systems  are far from those predicted in the infinitely heavy quark approximation. We checked by varying the coupling constants and masses of the light and $\bar s$ quarks over a wide range that this statement remains valid for reasonable $c$ quark masses. Therefore, we can conclude that the infinitely heavy quark approximation is inapplicable to the description of  $c\bar s$,  $c\bar u$, and $c\bar d$ systems. Therefore, any predictions of partial decay widths in these systems due to strong interactions based on the infinitely heavy quark approximation are questionable. 

Let's note that the masses of $D_{s0}$ and $D_{s1}$ mesons within our model are noticeably higher than their experimental values. This is possibly due to the fact that both of these states are close to the $DK$ and $D^*K$ thresholds, respectively. Therefore, $D_{s0}$ and $D_{s1}$ mesons may have significant admixture of molecular degrees of freedom \cite{Barnes:2003dj,Kolomeitsev:2003ac,Chen:2004dy,Guo:2006fu,Guo:2006rp}, which are not taken into account in our model.

However, it should be noted that, unlike all previous models, our model is qualitatively consistent with the currently observed properties of P-wave $D_s$ mesons. Specifically, the radiative decay of $D_{s1}(2535)$ to $D_s\gamma$ is not seen. The decay of $D_{s1}(2460)$ to $D_s\gamma$ is one of the main decays. An experimental test of the model could be to measure the decay probabilities of $D_{s1}(2535)$ and $D_{s1}(2460)$ into $D_s^*\gamma$. It is worth noting that the Belle II experiment recently discovered the decay of $D_{s0}(2317)$ into $D_s^*\gamma$ \cite{Belle-II:2025dzk}.

Previously, in Ref.~\cite{BM2025a} predictions have been obtained for  the radiative transition widths of $D_{s1}(2460)$ and $D_{s1}(2535)$ mesons. In Ref.~\cite{BM2025a}, only the first terms of the Hamiltonian  expansion in $\bm p^2/m^2$ are taken into account, in contrast to the present work. Naturally, the parameters in Ref.~\cite{BM2025a} differ noticeably from those used in our paper. However, the predictions in Ref.~\cite{BM2025a} for the radiative transition widths  and its hierarchy are in qualitative agreement with ours.

\section{Conclusion}
Using the relativistic potential model, a detailed analysis of the spectra and partial widths of radiative transitions in various systems of heavy mesons consisting either of a heavy quark and a heavy antiquark, or of a heavy quark and a light antiquark is performed. Although the number of parameters in the model is  small (the quark masses $m_c$, $m_b$, $m_s$, and $m_q$, the potential parameters $g$, $b$, and ${\cal C}_L$), it is possible to obtain qualitative agreement with all available numerous experimental data. In some cases, this agreement is achieved where it could not be achieved in all previous models. All this demonstrates the importance of taking relativistic effects into account. A remarkable property of the relativistic potential model is that the predictions for the meson masses and the partial widths of radiative transitions remain finite as the light quark mass tends to zero. The results we obtained may be useful in planning further experiments devoted to the study of heavy meson physics.

\section*{Acknowledgement}
The work of I.V. Obraztsov was supported by the Foundation for the Advancement of Theoretical Physics and Mathematics ”BASIS” under Grant No. 24-1-5-3-1.

\appendix
\section{Calculation of the amplitudes $t^{(0)}$, $t^{(S)}_{1,2}$, and $t^{(\Sigma)}_{1,2}$}\label{appA}
In this appendix, we present the explicit form of the amplitudes $t^{(0)}$, $t^{(S)}_{1,2}$, and $t^{(\Sigma)}_{1,2}$, calculated using variational wave functions (see Eqs.~\eqref{WF0} and \eqref{WF1}). These amplitudes, needed to calculate the transition widths, can be conveniently represented as
\begin{align}\label{ttau}
	&t^{(0)}=e_1\bar\tau^{(0)}-e_2\tau^{(0)},\, t^{(S)}_{1,2}=e_1\bar\tau^{(S)}_{1,2}-e_2\tau^{(S)}_{1,2},\, t^{(\Sigma)}_{1,2}=e_1\bar\tau^{(\Sigma)}_{1,2}+e_2\tau^{(\Sigma)}_{1,2}\,,
\end{align} 
where all amplitudes $\tau$ are obtained from the corresponding amplitudes $\bar\tau$ by replacing $m_1\leftrightarrow m_2$.
The amplitude $\bar\tau^{(0)}$, in accordance with Eq.~\eqref{G0}, is represented as a sum
\begin{align}\label{t0tot}
	&\bar\tau^{(0)}=\bar\tau^{(0)}_{1}+\bar\tau^{(0)}_{2}+\bar\tau^{(0)}_{3}\,.
\end{align} 
For the contribution $\bar\tau^{(0)}_{1}$ we have
\begin{align}\label{tau01}	
	&\bar\tau^{(0)}_{1}=-\dfrac{f^3\,\omega_0}{4\sqrt{\pi}\,\bar\omega}\int_0^\infty \dfrac{ds}{(1+s^2)^{5/2}}\exp\left[-\dfrac{m_1^2\bar\omega^2}{\omega_0^2\omega_1^2}s^2\right]\nonumber\\
	&\times\Bigg\{ \exp\left[-\dfrac{q_1^2(\omega_1^2+2\bar\omega^2s^2)}{4\bar\omega^2\omega_1^2(1+s^2)}\right]+\exp\left[-\dfrac{q_1^2(\omega_0^2+2\bar\omega^2s^2)}{4\bar\omega^2\omega_0^2(1+s^2)}\right]\Bigg\}\,,\nonumber\\
	&q_1=\dfrac{m_2 k}{(m_1+m_2)}\,,\quad \bar\omega=\sqrt{\dfrac{\omega_0^2+\omega_1^2}{2}}\,,\quad f=\dfrac{\sqrt{2\omega_0\omega_1}}{\bar\omega}\,,
\end{align} 
where $k$ is the photon energy. The formula \eqref{tau01} can be significantly simplified if we note that for $q_{1,2}\ll \bar\omega$ the exponents in the curly brackets can be replaced by unity for any $s$, while for $q_{1}\gg \bar\omega$ the main contribution to the integral comes from $s\ll 1$. Therefore, the curly bracket in Eq.~\eqref{tau01} can be replaced with good accuracy by $2\exp[-q_1^2/(4\bar\omega^2)]$ for any $s$. After this, we obtain
\begin{align}\label{t01a}	
	&\bar\tau^{(0)}_{1}=-\dfrac{f^3\,\omega_0}{3\,\bar\omega}{\cal G}\left(\beta_1/2\right)e^{-Q_1^2/4}\,,\nonumber\\
	&{\cal G}(x)=x[2x\,{\cal K}_0(x)+(1-2x)\,{\cal K}_1(x)]\,.
\end{align} 
The following notations are used here and below:
\begin{align}\label{def}	
	 &\beta_1=\dfrac{m_1^2\bar\omega^2}{\omega_0^2\omega_1^2}\,,\,\, \beta_2=\dfrac{m_2^2\bar\omega^2}{\omega_0^2\omega_1^2}\,,\,\, Q_1=\dfrac{m_2k}{M\bar\omega}\,,\quad Q_2=\dfrac{m_1k}{M\bar\omega}\,,\,\, M=(m_1+m_2)\,,\nonumber\\
	 &\alpha_{01}=\dfrac{\omega_0^2}{4\bar\omega^2}(y+\beta_1)\,,\,\, \alpha_{11}=\dfrac{\omega_1^2}{4\bar\omega^2}(y+\beta_1)\,,\,\,  \alpha_{02}=\dfrac{\omega_0^2}{4\bar\omega^2}(y+\beta_2)\,,\,\,\alpha_{12}=\dfrac{\omega_1^2}{4\bar\omega^2}(y+\beta_2)\,, \nonumber\\
	 &{\cal J}_0(x)=\dfrac{\sin x}{x},\,{\cal J}_1(x)=\dfrac{1}{x^3}[\sin x-x\cos x] ,\, {\cal J}_2(x)=\dfrac{1}{x^5}[(3-x^2)\sin x-3 x\cos x]\,,\nonumber\\
	 &{\cal J}_3(x)=\dfrac{1}{x^7}[3(5-2x^2)\sin x-x(15-x^2)\cos x]\,,
\end{align}
where $y$ is some integration variable (see below).

The contribution $\bar\tau^{(0)}_{2}$ reads
\begin{align}\label{t02}	
	&\bar\tau^{(0)}_{2}=\dfrac{gf\omega_1}{\sqrt{\pi}\,\bar\omega}\int_0^\infty e^{-y}dy \int_0^\infty re^{-r^2}dr\nonumber\\
	&\times \Bigg\{ \dfrac{\omega_0^2}{\bar\omega^2}yr^2  {\cal J}_1(Q_1r)\,{\cal K}_0(\alpha_{01}){\cal K}_0(\alpha_{12})+\alpha_{02}[{\cal K}_1(\alpha_{02})-{\cal K}_0(\alpha_{02})]\nonumber\\
	&\times\Bigg[{\cal J}_0(Q_1r){\cal K}_0(\alpha_{11})+{\cal J}_1(Q_1r)\Bigg({\cal K}_0(\alpha_{11})-\dfrac{\omega_1^2}{\bar\omega^2}yr^2[{\cal K}_0(\alpha_{11})+{\cal K}_1(\alpha_{11})]\Bigg)\Bigg]\Bigg\}\,.
\end{align} 
The contribution $\bar\tau^{(0)}_{3}$ has the form
\begin{align}\label{t03}	
	&\bar\tau^{(0)}_{3}=\dfrac{bf\,\omega_1}{2\sqrt{\pi}\,\bar\omega^3}
	\int_0^\infty ye^{-y}dy\,\int_0^\infty r^3e^{-r^2}dr\,\nonumber\\
	&\times\Bigg\{{\cal J}_0(Q_1r){\cal K}_0(\alpha_{01})[{\cal K}_0(\alpha_{11})+{\cal K}_1(\alpha_{11})] \nonumber\\
	&+\dfrac{yr^2}{4} {\cal J}_1(Q_1r)\Bigg[\dfrac{\omega_0^2}{\bar\omega^2}[{\cal K}_0(\alpha_{01})+{\cal K}_1(\alpha_{01})][{\cal K}_0(\alpha_{11})+{\cal K}_1(\alpha_{11})]\nonumber\\
	&-\dfrac{\omega_1^2}{\bar\omega^2}{\cal K}_0(\alpha_{01})\left[2{\cal K}_0(\alpha_{11})+\left(2+\dfrac{1}{\alpha_{11}}\right){\cal K}_1(\alpha_{11})\right]\Bigg]\Bigg\}\,.
\end{align} 
We have  
\begin{align}\label{tS1}	
	&\bar\tau^{(S)}_{1}=-\dfrac{2f\omega_1}{\sqrt{\pi}\,\bar\omega}\int_0^\infty e^{-y}dy\,\int_0^\infty re^{-r^2}dr\,{\cal J}_1(Q_1r)[{\cal Z}_1+{\cal Z}_2]\nonumber\\
	&+\dfrac{fk\omega_0}{4\bar\omega^2}\int_0^\infty \dfrac{dy}{(1+y)^{5/2}}e^{-\beta_1y}\,,\nonumber\\
	&{\cal Z}_1=g{\cal K}_0(\alpha_{11})\alpha_{01}[{\cal K}_1(\alpha_{01})-{\cal K}_0(\alpha_{01})]-\dfrac{yr^2b}{4\bar\omega^2}{\cal K}_0(\alpha_{01})[{\cal K}_0(\alpha_{11})+{\cal K}_1(\alpha_{11})]\,,\nonumber\\
	&{\cal Z}_2=g\Bigg[{\cal K}_0(\alpha_{12})\alpha_{01}[{\cal K}_1(\alpha_{01})-{\cal K}_0(\alpha_{01})]+{\cal K}_0(\alpha_{11})\alpha_{02}[{\cal K}_1(\alpha_{02})-{\cal K}_0(\alpha_{02})]\Bigg]\,,
\end{align} 
for amplitude $\bar\tau^{(S)}_{1}$, and  
\begin{align}\label{tS2}	
	&\bar\tau^{(S)}_{2}=-\dfrac{2f\omega_1}{\sqrt{\pi}\,\bar\omega}\int_0^\infty e^{-y}dy\,\int_0^\infty re^{-r^2}dr\,[{\cal J}_0(Q_1r)-3{\cal J}_1(Q_1r)][{\cal Z}_1+{\cal Z}_2]\nonumber\\
	&+\dfrac{fk\omega_0Q_1^2}{16\bar\omega^2}\int_0^\infty \dfrac{dy}{(1+y)^{7/2}}e^{-\beta_1y}\Bigg[\left(\dfrac{\omega_1^2}{\omega_0^2}-1\right)\left(1+\dfrac{\bar\omega^2}{\omega_1^2}y\right)+\dfrac{\bar\omega^4}{\omega_0^2\omega_1^2}y^2\Bigg]\nonumber\\
	&+\dfrac{2fk\omega_1Q_1}{3\bar\omega^2} e^{-Q_1^2/4}\Bigg[\left(1-\dfrac{\bar\omega^2}{\omega_1^2}\right){\cal G} (\beta_{1}/2)+\dfrac{3\bar\omega^2}{4\omega_1^2}\beta_1[{\cal K}_1(\beta_{1}/2)-{\cal K}_0(\beta_{1}/2)]\Bigg]\,,
\end{align} 
for the amplitude $\bar\tau^{(S)}_{2}$. The amplitudes $\bar\tau^{(\Sigma)}_{1}$ and $\bar\tau^{(\Sigma)}_{2}$ are obtained from the corresponding amplitudes $\bar\tau^{(S)}_{1}$ and $\bar\tau^{(S)}_{2}$ by replacing ${\cal Z}_2\to -{\cal Z}_2$.

\section{Radiative transition from a state with $L=0$ and $S_{tot}=1$ to a state with $L=0$ and $S_{tot}=0$}\label{appB}
The expression for the amplitude of radiative transition from a state with $L=0$ and $S_{tot}=1$ to a state with $L=0$ and $S_{tot}=0$ is obtained in Ref.~\cite{BM2025b}. For the convenience of readers, we present it in the terms of our work. The corresponding transition amplitude ${\cal\tau}(0,0,0|1,\mu,1)$ is represented as
\begin{align}\label{tau0}	
	&{\cal\tau}(0,0,0|1,\mu,1)=i([\bm e_\mu\times\bm n_k]\cdot\bm A_W^*)[e_1\bar\tau_0-e_2\tau_0]\,,\nonumber\\
	&\bar\tau_0=\dfrac{k}{2\omega_0}\exp\left(-\dfrac{q_1^2}{4\omega_0^2}\right)\gamma_{10}\left[{\cal K}_1(\gamma_{10}/2)-{\cal K}_0(\gamma_{10}/2)\right]\nonumber\\
	&+\dfrac{q_1}{2\omega_0\sqrt{\pi}}	\int_0^\infty e^{-y}dy\,\int_0^\infty r e^{-r^2}dr\,{\cal J}_1(q_1r/\omega_0)\nonumber\\
	&\times\Bigg[g{\cal K}_0(z_{10})[{\cal K}_0(z_{10})-2{\cal K}_0(z_{20})]-\dfrac{br^2}{4\omega_0^2}y[{\cal K}_0(z_{10})+{\cal K}_1(z_{10})]^2\Bigg]\,,
\end{align}
where $\gamma_{10}$, $\gamma_{20}$, $z_{10}$, and $z_{20}$ are defined in Eqs.~\eqref{E0} and \eqref{DE00}. The amplitude $\tau_0$ is obtained from the amplitude $\bar\tau_0$ by replacing $q_1\to q_2$, $\gamma_{10}\to\gamma_{20}$, and $z_{10}\leftrightarrow z_{20}$.
 
\section{Radiative transition from states with $J=1$ and $L=1$ to a state with $J=0$ and $L=1$}\label{appC}
Formulas for the amplitudes and partial widths of the radiative transition from states with $J=1$, $L=1$ to a state with $J=0$, $L=1$ are given in Eqs.~\eqref{T1T0} and \eqref{Gam11Gam01}. The coefficients in these formulas have the form
\begin{align}\label{TTau}
	&T^{(0)}=e_1\bar{\cal M}^{(0)}+e_2{\cal M}^{(0)},\, T^{(S)}=e_1\bar{\cal M}^{(S)}+e_2{\cal M}^{(S)},\, T^{(\Sigma)}=e_1\bar{\cal M}^{(\Sigma)}-e_2{\cal M}^{(\Sigma)}\,,
\end{align} 
where all amplitudes ${\cal M}$ are obtained from the corresponding amplitudes $\bar{\cal M}$ by replacing $m_1\leftrightarrow m_2$. We have for $\bar{\cal M}^{(0)}$
\begin{align}\label{T0}	
	&\bar{\cal M}^{(0)}=\dfrac{2q_1}{3\sqrt{6}\omega_1}\exp\left(-\dfrac{q_1^2}{4\omega_1^2}\right)\gamma_{11}\Big[\gamma_{11}{\cal K}_0(\gamma_{11}/2)+(1-\gamma_{11}){\cal K}_1(\gamma_{11}/2)\Big]\nonumber\\
	&-\dfrac{q_1}{2\omega_1\sqrt{6\pi}}\int_0^\infty ye^{-y}dy\,\int_0^\infty r^3 e^{-r^2}dr\,{\cal J}_1(q_1r/\omega_1)\nonumber\\
	&\times\Bigg\{4g{\cal K}_0(z_{11}){\cal K}_0(z_{21})+\dfrac{byr^2}{\omega_1^2}\,\Big[{\cal K}_0(z_{11})+{\cal K}_1(z_{11})\Big]^2\Bigg\}\,.
\end{align}
Here $\gamma_{11}$, $\gamma_{21}$, $z_{11}$, and $z_{21}$ are defined in Eqs.~\eqref{E1} and \eqref{DE10}. The expression for $\bar{\cal M}^{(S)}$ reads
\begin{align}\label{TS}	
	&\bar{\cal M}^{(S)}=-\dfrac{2k}{3\sqrt{6}\omega_1}\exp\left(-\dfrac{q_1^2}{4\omega_1^2}\right)\gamma_{11}\Bigg\{ \Big[\gamma_{11}{\cal K}_0(\gamma_{11}/2)+(1-\gamma_{11}){\cal K}_1(\gamma_{11}/2)\Big]\nonumber\\
	&+\dfrac{q_1^2}{20\omega_1^2}\Big[-\gamma_{11}(2\gamma_{11}+9){\cal K}_0(\gamma_{11}/2)+(2\gamma_{11}^2+7\gamma_{11}-6){\cal K}_1(\gamma_{11}/2)\Big]\Bigg\}\nonumber\\
	&-\dfrac{q_1}{8\sqrt{6\pi}\omega_1}\int_0^\infty ye^{-y}dy\,\int_0^\infty r^3 e^{-r^2}dr\Bigg\{40g{\cal J}_2(q_1r/\omega_1){\cal K}_0(z_{11})\Big[{\cal K}_0(z_{11})+2{\cal K}_0(z_{21})\Big]\nonumber\\
	&-\dfrac{gyr^2q_1^2}{\omega_1^2}{\cal J}_3(q_1r/\omega_1)\Big[{\cal K}_0(z_{11})+{\cal K}_1(z_{11})\Big]\Big[{\cal K}_0(z_{11})+{\cal K}_1(z_{11})+2{\cal K}_0(z_{21})+2{\cal K}_1(z_{21})\Big]\nonumber\\
	&-\dfrac{byr^2}{\omega_1^2}\Bigg[5{\cal J}_2(q_1r/\omega_1)\Big[{\cal K}_0(z_{11})+{\cal K}_1(z_{11})\Big]^2-\dfrac{yr^2q_1^2}{12\omega_1^2z_{11}^2}{\cal J}_3(q_1r/\omega_1)\nonumber\\
	&\times\Big[(1+2z_{11}){\cal K}_1(z_{11})+2z_{11}{\cal K}_0(z_{11})\Big]^2\Bigg]	\Bigg\}-\dfrac{kq_1}{8\sqrt{6}\omega_1^2}\exp\left(-\dfrac{q_1^2}{4\omega_1^2}\right)\int_0^\infty\dfrac{\exp(-\gamma_{11}y)\,dy}{(1+y)^{5/2}}\,.
\end{align}
For $\bar{\cal M}^{(\Sigma)}$ we have
\begin{align}\label{TSig}	
	&\bar{\cal M}^{(\Sigma)}=-\dfrac{k}{3\sqrt{3}\omega_1}\exp\left(-\dfrac{q_1^2}{4\omega_1^2}\right)\gamma_{11} \Big[\gamma_{11}{\cal K}_0(\gamma_{11}/2)+(1-\gamma_{11}){\cal K}_1(\gamma_{11}/2)\Big]\nonumber\\
	&-\dfrac{kq_1}{8\sqrt{3}\omega_1^2}\exp\left(-\dfrac{q_1^2}{4\omega_1^2}\right)\int_0^\infty\dfrac{\exp(-\gamma_{11}y)\,dy}{(1+y)^{5/2}}\,.
\end{align}
For definitions of ${\cal J}_0(x)$, ${\cal J}_1(x)$, ${\cal J}_2(x)$, and ${\cal J}_3(x)$, see Eq.~\eqref{def}.

Our predictions for the partial widths of radiative transitions from the states with $J=1$ and $L=1$ to a state with $J=0$ and $L=1$ for various systems are given in the Table~\ref{J1J0L1tab}. Comparison of the partial widths for the upper and lower states with $J=1$ shows that taking mixing into account is very important when calculating transition probabilities.
\begin{table}[!h]
	\caption{\label{J1J0L1tab} Prediction for partial radiative transition widths (in \mbox{eV})  from states with $J=1$ and $L=1$ to a state with $J=0$ and $L=1$.}
	\begin{center}
		\begin{tabular}{lcccc} 
			&$\Gamma(h_b\to \chi_{b0}\gamma)$  &$\Gamma(h_c\to \chi_{c0}\gamma)$ &$\Gamma(B_{c1}\to B_{c0}\gamma)$ &$\Gamma(B_{c1}'\to B_{c0}\gamma)$\\
			\hline
			
			Predictions &0.17 &95.2 &0.46 &9.23\\ 
			\hline
			
			&$\Gamma(D_{s1}\to D_{s0}\gamma)$ &$\Gamma(D_{s1}'\to D_{s0}\gamma)$ &$\Gamma(B_{s1}\to B_{s0}\gamma)$ &$\Gamma(B_{s1}'\to B_{s0}\gamma)$\\
			\hline
			
			Predictions &218.5 &10.1 &0.002 &15.71\\ 
			\hline
			
			&$\Gamma(D^{(+)}_{1}\to D^{(+)}_{0}\gamma)$ &$\Gamma(D^{(+)}_{1}{}'\to D^{(+)}_{0}\gamma)$ &$\Gamma(D^{(0)}_{1}\to D^{(0)}_{0}\gamma)$ &$\Gamma(D^{(0)}_{1}{}'\to D^{(0)}_{0}\gamma)$\\
			\hline
			
			Predictions &493 &0.08 &465 &944\\ 
			\hline
			
			&$\Gamma(B^{(-)}_{1}\to B^{(-)}_{0}\gamma)$ &$\Gamma(B^{(-)}_{1}{}'\to B^{(-)}_{0}\gamma)$ &$\Gamma(B^{(0)}_{1}\to B^{(0)}_{0}\gamma)$ &$\Gamma(B^{(0)}_{1}{}'\to B^{(0)}_{0}\gamma)$\\
			\hline
			
			Predictions &18.2 &193 &1.4 &53.8\\ 
			\hline
		\end{tabular}
	\end{center}
\end{table}


\begin{thebibliography}{99}
	\bibitem{Bykov:1984}
	{A.~A.~Bykov, I.~M.~Dremin and A.~V.~Leonidov},
	\href{https://doi.org/10.1070/PU1984v027n05ABEH004291}{Sov. Phys. Usp. \textbf{27}, 321 (1984)}.
	
	\bibitem{GI1985}
	{S.~Godfrey and N.~Isgur},
	\href{https://doi.org/10.1103/PhysRevD.32.189}{Phys. Rev. D \textbf{32}, 189 (1985)}.
	
	\bibitem{Breit:1929zz}
	{G.~Breit},
	\href{https://doi.org/10.1103/PhysRev.34.553}{Phys. Rev. \textbf{34}, 553 (1929)}.
	
	\bibitem{Bondar:2025gsh}
	{A.~E.~Bondar},
	\href{https://doi.org/10.1134/S0021364024605037}{JETP Lett. \textbf{121}, 231 (2025)}.
	
	\bibitem{BM2025a}
	{A.~E.~Bondar and A.~I.~Milstein},
	\href{https://doi.org/10.1103/64y1-qyvt}{Phys. Rev. D \textbf{111}, 114019 (2025)}.
	
	\bibitem{BM2025b}
	{A.~E.~Bondar and A.~I.~Milstein}, 
	\href{https://doi.org/10.1103/gnyk-r31j}{Phys. Rev. D \textbf{112}, 054037 (2025)}.
	
	\bibitem{Pilkuhn:1979ps}
	{H.~M.~Pilkuhn},
	\href{https://doi.org/10.1007/978-3-642-88079-7}{Springer Berlin, Heidelberg (1979)}.
	
	\bibitem{Lee:2001wwa}
	{R.~N.~Lee, A.~I.~Milstein and M.~Schumacher},
	\href{https://doi.org/10.1103/PhysRevA.64.032507}{Phys. Rev. A \textbf{64}, 032507 (2001)}.
	
	\bibitem{LLQM} 
	{L.~D.~Landau and E.~M.~Lifshitz},
	\href{https://doi.org/10.1016/C2013-0-02793-4}{Pergamon Press, Oxford (1977)}.
	
	\bibitem{Matsuki:2010zy}
	{T.~Matsuki, T.~Morii and K.~Seo},
	\href{https://doi.org/10.1143/PTP.124.285}{Prog. Theor. Phys. \textbf{124}, 285 (2010)}.
	
	\bibitem{Navas:2024ynf}
	S.~Navas \textit{et al.} (Particle Data Group),
	\href{https://doi.org/10.1103/PhysRevD.110.030001}{Phys. Rev. D \textbf{110}, 030001 (2024)}.
	
	\bibitem{Ebert:2002pp}
	{D.~Ebert, R.~N.~Faustov and V.~O.~Galkin},
	\href{https://doi.org/10.1103/PhysRevD.67.014027}{Phys. Rev. D \textbf{67}, 014027 (2003)}.
	
	\bibitem{Barnes:2005pb}
	{T.~Barnes, S.~Godfrey and E.~S.~Swanson},
	\href{https://doi.org/10.1103/PhysRevD.72.054026}{Phys. Rev. D \textbf{72}, 054026 (2005)}.
	
	\bibitem{Radford:2007vd}
	{S.~F.~Radford and W.~W.~Repko},
	\href{https://doi.org/10.1103/PhysRevD.75.074031}{Phys. Rev. D \textbf{75}, 074031 (2007)}.
	
	\bibitem{Dudek:2009kk}
	{J.~J.~Dudek, R.~Edwards and C.~E.~Thomas},
	\href{https://doi.org/10.1103/PhysRevD.79.094504}{Phys. Rev. D \textbf{79}, 094504 (2009)}.
	
	\bibitem{Li:2009zu}
	{B.~Q.~Li and K.~T.~Chao},
	\href{https://doi.org/10.1103/PhysRevD.79.094004}{Phys. Rev. D \textbf{79}, 094004 (2009)}.
	
	\bibitem{Godfrey:2015dia}
	{S.~Godfrey and K.~Moats},
	\href{https://doi.org/10.1103/PhysRevD.92.054034}{Phys. Rev. D \textbf{92}, 054034 (2015)}.
	
	\bibitem{Li:2009nr}
	{B.~Q.~Li and K.~T.~Chao},
	\href{https://doi.org/10.1088/0253-6102/52/4/20}{Commun. Theor. Phys. \textbf{52}, 653 (2009)}.
	
	\bibitem{li:2021hss}
	{Q.~li, R.~H.~Ni and X.~H.~Zhong},
	\href{https://doi.org/10.1103/PhysRevD.103.116010}{Phys. Rev. D \textbf{103}, 116010 (2021)}.
	
	\bibitem{Godfrey:2016nwn}
	{S.~Godfrey, K.~Moats and E.~S.~Swanson},
	\href{https://doi.org/10.1103/PhysRevD.94.054025}{Phys. Rev. D \textbf{94}, 054025 (2016)}.
	
	\bibitem{Lu:2016bbk}
	{Q.~F.~L\"u, T.~T.~Pan, Y.~Y.~Wang, E.~Wang and D.~M.~Li},
	\href{https://doi.org/10.1103/PhysRevD.94.074012}{Phys. Rev. D \textbf{94}, 074012 (2016)}.
	
	\bibitem{Asghar:2018tha}
	{I.~Asghar, B.~Masud, E.~S.~Swanson, F.~Akram and M.~Atif Sultan},
	\href{https://doi.org/10.1140/epja/i2018-12558-6}{Eur. Phys. J. A \textbf{54}, 127 (2018)}.
	
	\bibitem{Colangelo:2012xi}
	{P.~Colangelo, F.~De Fazio, F.~Giannuzzi and S.~Nicotri},
	\href{https://doi.org/10.1103/PhysRevD.86.054024}{Phys. Rev. D \textbf{86}, 054024 (2012)}.
	
	\bibitem{Akan:2025nej}
	{T.~Akan},
	\href{https://doi.org/10.48550/arXiv.2511.10986}{arXiv:hep-ph/2511.10986}.
	
	\bibitem{Soni:2017wvy}
	{N.~R.~Soni, B.~R.~Joshi, R.~P.~Shah, H.~R.~Chauhan and J.~N.~Pandya},
	\href{https://doi.org/10.1140/epjc/s10052-018-6068-6}{Eur. Phys. J. C \textbf{78},  592 (2018)}.
	
	\bibitem{Mutuk:2018erw}
	{H.~Mutuk},
	\href{https://doi.org/10.1155/2019/3105373}{Adv. High Energy Phys. \textbf{2019}, 3105373 (2019)}.
	
	\bibitem{Devlani:2014nda}
	{N.~Devlani, V.~Kher and A.~K.~Rai},
	\href{https://doi.org/10.1140/epja/i2014-14154-2}{Eur. Phys. J. A \textbf{50},  154 (2014)}.
	
	\bibitem{Ebert:2011jc}
	{D.~Ebert, R.~N.~Faustov and V.~O.~Galkin},
	\href{https://doi.org/10.1140/epjc/s10052-011-1825-9}{Eur. Phys. J. C \textbf{71}, 1825 (2011)}.
	
	\bibitem{Godfrey:2004ya}
	{S.~Godfrey},
	\href{https://doi.org/10.1103/PhysRevD.70.054017}{Phys. Rev. D \textbf{70}, 054017 (2004)}.
	
	\bibitem{Eichten:1994gt}
	{E.~J.~Eichten and C.~Quigg},
	\href{https://doi.org/10.1103/PhysRevD.49.5845}{Phys. Rev. D \textbf{49}, 5845 (1994)}.
	
	\bibitem{Monteiro:2016rzi}
	{A.~P.~Monteiro, M.~Bhat and K.~B.~Vijaya Kumar},
	\href{https://doi.org/10.1103/PhysRevD.95.054016}{Phys. Rev. D \textbf{95}, 054016 (2017)}.
	
	\bibitem{Gershtein:1994dxw}
	{S.~S.~Gershtein, V.~V.~Kiselev, A.~K.~Likhoded and A.~V.~Tkabladze}
	\href{https://doi.org/10.1103/PhysRevD.51.3613}{Phys. Rev. D \textbf{51}, 3613 (1995)}.
	
	\bibitem{Li:2023wgq}
	{X.~J.~Li, Y.~S.~Li, F.~L.~Wang and X.~Liu},
	\href{https://doi.org/10.1140/epjc/s10052-023-12237-9}{Eur. Phys. J. C \textbf{83}, 1080 (2023)}.
	
	\bibitem{Li:2022bre}
	{T.~y.~Li, L.~Tang, Z.~y.~Fang, C.~h.~Wang, C.~q.~Pang and X.~Liu},
	\href{https://doi.org/10.1103/PhysRevD.108.034019}{Phys. Rev. D \textbf{108}, 034019 (2023)}.
	
	\bibitem{Li:2019tbn}
	{Q.~Li, M.~S.~Liu, L.~S.~Lu, Q.~F.~L{\"u}, L.~C.~Gui and X.~H.~Zhong},
	\href{https://doi.org/10.1103/PhysRevD.99.096020}{Phys. Rev. D \textbf{99}, 096020 (2019)}.
	
	\bibitem{Eichten:2019gig}
	{E.~J.~Eichten and C.~Quigg},
	\href{https://doi.org/10.1103/PhysRevD.99.054025}{Phys. Rev. D \textbf{99}, 054025 (2019)}.
	
	\bibitem{LHCb:2025uce}
	{R.~Aaij \textit{et al.} (LHCb Collaboration)},
	\href{https://doi.org/10.1103/fc8j-tb8k}{Phys. Rev. Lett. \textbf{135}, 231902 (2025)}.
	
	\bibitem{LHCb:2025ubr}
	{R.~Aaij \textit{et al.} (LHCb Collaboration)},
	\href{https://doi.org/10.1103/1d49-q8h4}{Phys. Rev. D \textbf{112}, 112003 (2025)}.
	
	\bibitem{Li:2022vby}
	{J.~L.~Li and D.~Y.~Chen},
	\href{https://doi.org//10.1088/1674-1137/ac600c}{Chin. Phys. C \textbf{46}, 073106 (2022)}.
	
	\bibitem{Godfrey:2015dva}
	{S.~Godfrey and K.~Moats},
	\href{https://doi.org//10.1103/PhysRevD.93.034035}{Phys. Rev. D \textbf{93}, 034035 (2016)}.
	
	\bibitem{Close:2005se}
	{F.~E.~Close and E.~S.~Swanson},
	\href{https://doi.org/10.1103/PhysRevD.72.094004}{Phys. Rev. D \textbf{72}, 094004 (2005)}.
	
	\bibitem{Colangelo:1993zq}
	{P.~Colangelo, F.~De Fazio and G.~Nardulli},
	\href{https://doi.org/10.1016/0370-2693(93)91043-M}{Phys. Lett. B \textbf{316}, 555 (1993)}.
	
	\bibitem{Korner:1992pz}
	{J.~G.~Korner, D.~Pirjol and K.~Schilcher},
	\href{https://doi.org/10.1103/PhysRevD.47.3955}{Phys. Rev. D \textbf{47}, 3955 (1993)}.
	
	\bibitem{Godfrey:2005ww}
	{S.~Godfrey},
	\href{https://doi.org/10.1103/PhysRevD.72.054029}{Phys. Rev. D \textbf{72}, 054029 (2005)}.
	
	\bibitem{Goity:2000dk}
	{J.~L.~Goity and W.~Roberts},
	\href{https://doi.org/10.1103/PhysRevD.64.094007}{Phys. Rev. D \textbf{64}, 094007 (2001)}.
	
	\bibitem{Green:2016occ}
	{N.~Green, W.~W.~Repko and S.~F.~Radford},
	\href{https://doi.org/10.1016/j.nuclphysa.2016.11.006}{Nucl. Phys. A \textbf{958}, 71 (2017)}.
	
	\bibitem{Radford:2009bs}
	{S.~F.~Radford, W.~W.~Repko and M.~J.~Saelim},
	\href{https://doi.org/10.1103/PhysRevD.80.034012}{Phys. Rev. D \textbf{80}, 034012 (2009)}.
	
	\bibitem{Chen:2020jku}
	{S.~F.~Chen, J.~Liu, H.~Q.~Zhou and D.~Y.~Chen},
	\href{https://doi.org/10.1140/epjc/s10052-020-7852-7}{Eur. Phys. J. C \textbf{80}, 290 (2020)}.
	
	\bibitem{Tran:2023hrn}
	{C.~T.~Tran, M.~A.~Ivanov, P.~Santorelli and Q.~C.~Vo},
	\href{https://doi.org/10.1088/1674-1137/ad102c}{Chin. Phys. C \textbf{48}, 023103 (2024)}.
	
	\bibitem{Pullin:2021ebn}
	{B.~Pullin and R.~Zwicky},
	\href{https://doi.org/10.1007/JHEP09(2021)023}{JHEP \textbf{09}, 023 (2021)}.
	
	\bibitem{Becirevic:2009xp}
	{D.~Becirevic and B.~Haas},
	\href{https://doi.org/10.1140/epjc/s10052-011-1734-y}{Eur. Phys. J. C \textbf{71}, 1734 (2011)}.
	
	\bibitem{Donald:2013sra}
	{G.~C.~Donald \textit{et al.} (HPQCD Collaboration)},
	\href{https://doi.org/10.1103/PhysRevLett.112.212002}{Phys. Rev. Lett. \textbf{112}, 212002 (2014)}.
	
	\bibitem{Deng:2013uca}
	{H.~B.~Deng, X.~L.~Chen, and W.~Z.~Deng},
	\href{https://doi.org/10.1088/1674-1137/38/1/013103}{Chin. Phys. C \textbf{38}, 013103 (2014)}.
	
	\bibitem{Ebert:2002xz}
	{D.~Ebert, R.~N.~Faustov, and V.~O.~Galkin},
	\href{https://doi.org/10.1016/S0370-2693(02)01939-1}{Phys. Lett. B \textbf{537}, 241 (2002)}.
	
	\bibitem{Choi:2007se}
	{H.~M.~Choi},
	\href{https://doi.org/10.1103/PhysRevD.75.073016}{Phys. Rev. D \textbf{75}, 073016 (2007)}.
	
	\bibitem{Zhu:1996qy}
	{S.~L.~Zhu, Z.~S.~Yang and W.~Y.~P.~Hwang},
	\href{https://doi.org/10.1142/S0217732397003150}{Mod. Phys. Lett. A \textbf{12}, 3027 (1997)}.
	
	\bibitem{Aliev:1994nq}
	{T.~M.~Aliev, E.~Iltan and N.~K.~Pak},
	\href{https://doi.org/10.1016/0370-2693(94)90606-8}{Phys. Lett. B \textbf{334}, 169 (1994)}.

    \bibitem{Barnes:2003dj}
    {T.~Barnes, F.~E.~Close and H.~J.~Lipkin},
    \href{https://doi.org/10.1103/PhysRevD.68.054006}{Phys. Rev. D \textbf{68}, 054006 (2003)}.

    \bibitem{Kolomeitsev:2003ac}
    {E.~E.~Kolomeitsev and M.~F.~M.~Lutz},
    \href{https://doi.org/10.1016/j.physletb.2003.10.118}{Phys. Lett. B \textbf{582}, 39 (2004)}.

    \bibitem{Chen:2004dy}
    {Y.~Q.~Chen and X.~Q.~Li},
    \href{https://doi.org/10.1103/PhysRevLett.93.232001}{Phys. Rev. Lett. \textbf{93}, 232001 (2004)}.

    \bibitem{Guo:2006fu}
    {F.~K.~Guo, P.~N.~Shen, H.~C.~Chiang, R.~G.~Ping and B.~S.~Zou},
    \href{https://doi.org/10.1016/j.physletb.2006.08.064}{Phys. Lett. B \textbf{641}, 278 (2006)}.

    \bibitem{Guo:2006rp}
    {F.~K.~Guo, P.~N.~Shen and H.~C.~Chiang},
    \href{https://doi.org/10.1016/j.physletb.2007.01.050}{Phys. Lett. B \textbf{647}, 133 (2007)}.

    \bibitem{Belle-II:2025dzk}
    {M.~Abumusabh \textit{et al.} (Belle-II Collaboration)},
    \href{https://doi.org/10.48550/arXiv.2510.27174}{arXiv:hep-ex/2510.27174}.
\end{thebibliography}
\end{document}